\documentclass[12pt]{article}
\pdfoutput=1
\usepackage{putex}

\usepackage{graphicx}
\usepackage{caption}
\usepackage{amsmath}
\usepackage{amssymb}
\usepackage{array}
\usepackage{multirow}
\usepackage{mathtools}
\usepackage{comment}
\usepackage{subcaption}
\usepackage{epstopdf}
\usepackage{enumerate}
\usepackage{cite}
\usepackage{youngtab}
\usepackage{tensor}
\usepackage{slashed}
\usepackage[aligntableaux=center]{ytableau}
\usepackage[utf8]{inputenc}
\usepackage{rotating}
\usepackage{bigfoot}
\usepackage[
      colorlinks=true,
      linkcolor=blue,
      urlcolor=blue,
      filecolor=black,
      citecolor=red,
      linktocpage=true
      ]{hyperref}

\newcommand{\abs}[1]{\left\lvert #1 \right\rvert}

\newcommand {\be} {\begin {equation}}
\newcommand {\ee} {\end {equation}}

\newcommand {\bes} {\begin {equation*}}
\newcommand {\ees} {\end {equation*}}
\newcommand{\es}[2] {\begin{equation} \label{#1} \begin{split} #2 \end{split} \end{equation}}

\newcommand{\Z}{\mathbb{Z}}

\newcommand{\cG}{{\mathcal G}}

\newcommand{\cN}{{\mathcal N}}
\newcommand{\cO}{{\mathcal O}}

\newcommand   \zb  {\bar{z}}

\newcommand{\bea}{\begin{equation}\begin{aligned}}
\newcommand{\eea}[1]{\label{#1}\end{aligned}\end{equation}}
\newcommand{\beq}{\begin{equation}}
\newcommand{\eeq}{\end{equation}}

\def\ie{\begin{equation}\begin{aligned}}
\def\fe{\end{aligned}\end{equation}}

\numberwithin{equation}{section}
\def\<{\langle}
\def\>{\rangle}

\newcommand{\balpha}{\bar{\alpha}}
\newcommand{\bbeta}{\bar{\beta}}
\newcommand{\by}{\bar{y}}
\newcommand{\bfp}{\mathbf{p}}
\newcommand{\bfq}{\mathbf{q}}
\newcommand{\bfg}{\mathbf{g}}
\newcommand{\bft}{\mathbf{2}}
\newcommand{\bfo}{\mathbf{O}}
\newcommand{\co}{\mathcal{O}}

\makeatletter
\DeclareFontFamily{OMX}{MnSymbolE}{}
\DeclareSymbolFont{MnLargeSymbols}{OMX}{MnSymbolE}{m}{n}
\SetSymbolFont{MnLargeSymbols}{bold}{OMX}{MnSymbolE}{b}{n}
\DeclareFontShape{OMX}{MnSymbolE}{m}{n}{
    <-6>  MnSymbolE5
   <6-7>  MnSymbolE6
   <7-8>  MnSymbolE7
   <8-9>  MnSymbolE8
   <9-10> MnSymbolE9
  <10-12> MnSymbolE10
  <12->   MnSymbolE12
}{}
\DeclareFontShape{OMX}{MnSymbolE}{b}{n}{
    <-6>  MnSymbolE-Bold5
   <6-7>  MnSymbolE-Bold6
   <7-8>  MnSymbolE-Bold7
   <8-9>  MnSymbolE-Bold8
   <9-10> MnSymbolE-Bold9
  <10-12> MnSymbolE-Bold10
  <12->   MnSymbolE-Bold12
}{}
\let\llangle\@undefined
\let\rrangle\@undefined
\DeclareMathDelimiter{\llangle}{\mathopen}%
                     {MnLargeSymbols}{'164}{MnLargeSymbols}{'164}
\DeclareMathDelimiter{\rrangle}{\mathclose}%
                     {MnLargeSymbols}{'171}{MnLargeSymbols}{'171}
\newcommand{\exmm}[1]{\llangle #1 \rrangle}

\begin{document}

\preprint{}

\institution{imperial}{Blackett Laboratory, Imperial College, Prince Consort Road, London, SW7 2AZ, U.K.}
\institution{stony}{Simons Center for Geometry and Physics, SUNY, Stony Brook, NY 11794, USA}

\title{Modular invariant gluon-graviton scattering\\ in AdS at one loop}

\authors{Shai M.~Chester,\worksat{\imperial} Pietro Ferrero\worksat{\stony} and Daniele R. Pavarini\worksat{\imperial}}

\abstract{
We consider mixed gluon-graviton scattering in Type IIB string theory on AdS$_5\times S^5/\mathbb{Z}_2$ in the presence of D7 branes, which is dual to a mixed correlator of the $SO(8)$ and $SU(2)_L$ flavor multiplets of a certain 4d $\mathcal{N}=2$ $USp(2N)$ gauge theory with complexified coupling $\tau$. We compute this holographic correlator in the large $N$ and finite $\tau$ expansion using constraints from derivatives of the mass deformed sphere free energy, which we compute using supersymmetric localization at large $N$ and finite $\tau$ in terms of modular invariant non-holomorphic Eisenstein series. In particular, we combine this constraint with the known flat space limit to fix the $R^2F^2$ higher derivative correction to the correlator
in terms of the weight one non-holomorphic Eisenstein series, and also to fix the logarithmic threshold. We also compute the one-loop correction to the correlator, and match it to the expected flat space limit result.
}
\date{}

\maketitle

\tableofcontents
\newpage

\section{Introduction}\label{sec:intro}

The AdS/CFT correspondence relates particles scattering in ($d+1$)-dimensional anti-de Sitter (AdS) space to correlation functions in the dual conformal field theory (CFT) in $d$ dimensions \cite{Maldacena:1997re}. In the large $N$ limit, these holographic correlators can be computed from the bulk supergravity using Witten diagrams, which was done {\it e.g.} for graviton scattering in $AdS_5\times S^5$ in \cite{Arutyunov:2000py}. A much more efficient method was developed in \cite{Rastelli:2017udc}, which used the constraints of crossing symmetry, superconformal symmetry, analyticity, and the flat space limit to completely fix the correlators of all Kaluza-Klein (KK) modes that arise from decomposing the 10d graviton on $AdS_5\times S^5$. This analytic bootstrap method was soon generalized to graviton scattering in other dimensions with maximal supersymmetry ({\it i.e.} $d=3,6$) \cite{Rastelli:2017ymc,Zhou:2017zaw,Alday:2020dtb}, as well as to gluons scattering on branes (or M-theory singularities) in setups with half-maximal supersymmetry in $d=3,4,5,6$ \cite{Zhou:2018ofp,Alday:2021odx}. Higher derivative corrections to supergravity were also computed in these cases using constraints from supersymmetric localization \cite{Binder:2019jwn,Binder:2021cif,Alday:2021ymb,Binder:2018yvd,Binder:2019mpb,Chester:2023qwo,Chester:2021aun,Chester:2020dja,Chester:2019pvm,Chester:2019jas,Alday:2021vfb,Alday:2022rly,Chester:2024bij,Chester:2024esn,Alday:2024yax,Chester:2023ehi,Alday:2023pet,Dempsey:2024vkf,Billo:2023ncz,Brown:2024yvt,Alday:2024srr,Cavaglia:2023mmu,Cavaglia:2022qpg,Caron-Huot:2024tzr,Paul:2023zyr,Glew:2023wik}. In the case of graviton scattering in type IIB string theory on $AdS_5\times S^5$, the holographic correlator was modular invariant under the complexified string coupling $\tau_s=\chi+i/g_s$ due to $SL(2,\mathbb{Z})$ duality symmetry, as imposed by the localization constraints \cite{Chester:2019jas,Chester:2020vyz,Alday:2021vfb}.

The case of mixed gluon-graviton scattering was first considered for gluons on an M-theory orbifold singularity with an $AdS_4\times S^3$ fixed point locus, which is dual to a certain 3d CFT \cite{Chester:2023qwo}. Constraints from supersymmetric localization combined with the known flat space limit were used to show the vanishing of the first two higher derivative corrections, denoted as $RF^2$ and $R^2F^2$. In this paper, we will consider mixed gluon-graviton scattering in type IIB string theory on an $AdS_5\times S^5/\mathbb{Z}_2$ orientifold, which arises in the large $N$ limit of $N$ D3 branes probing a $D_4$ F-theory singularity \cite{Sen:1996vd,Banks:1996nj}, and has an $AdS_5\times S^3$ fixed point locus. The dual CFT is a 4d $\mathcal{N}=2$ $USp(2N)$ gauge theory with four fundamental hypermultiplets, one antisymmetric hypermultiplet, and a $SO(8)\times SU(2)_L$ flavor symmetry. 
This theory is a conformal manifold with one complex parameter $\tau$,\footnote{This is the IR $\tau$, which is related to the usual UV $\tau_\text{UV} = \frac{\theta}{2\pi}+i\frac{4\pi}{g_\text{YM}^2}$ as $e^{4\pi i \tau_\text{UV}}=16\frac{\theta_2(\tau/2)^4}{\theta_3(\tau/2)^4}$ \cite{Hollands:2010xa,Douglas:1996js}.} which is related to bulk parameters as
\es{adscft}{
\frac{L^4}{\ell_s^4}=8\pi g_s N\,,\qquad \tau_s\equiv\frac{\chi}{2\pi}+\frac{i}{g_s}=\tau\,,
}
where $L$ is the AdS radius and $\ell_s$ is the string length. Previous work computed gluon scattering in this setup, where the gluons are dual to the flavor multiplet of the $SO(8)$ flavor symmetry. In this work, the graviton modes we consider are dual to flavor multiplets of the $SU(2)_L$ flavor symmetry\footnote{Decomposing the 10d graviton on $AdS_5\times S^5/\mathbb{Z}_2$ leads to several towers of KK modes, unlike the $AdS_5\times S^5$ case that just has a single tower.}, which arises upon decomposing $SU(4)\to SU(2)_R\times U(1)_R\times SU(2)_L$ when breaking the $\mathcal{N}=4$ superconformal symmetry of $AdS_5\times S^5$ down to the $\mathcal{N}=2$ symmetry of the orientifold. The large $N$ expansion of the graviton-gluon holographic correlator $G$ is fixed by the analytic bootstrap to take the form: 
\es{introG}{
G=G^{(0)}+\frac{G^{(R)}}{N^2}+\frac{G^{(RF^2)}}{N^{5/2}}+\frac{1}{N^3}\big[G^{(R|F^2)}+G^{(R^2F^2)}(\tau)+G^{(\log)}\log N\big]+O(N^{-4})\,,
}
where the powers of $N$ follow from the AdS/CFT dictionary \eqref{adscft}. Here, $G^{(0)}$ is the disconnected term, $G^{(R)}$ arises from tree level graviton exchange, $G^{(RF^2)}$ and $G^{(R^2F^2)}$ are contact terms due to the respective higher derivative correction, $G^{(\log)}$ is a contact term due to the logarithmic threshold, and $G^{(R|F^2)}$ is a one-loop term with one graviton $R$ and one gluon $F^2$ vertex.

We fix the tree level $G^{(R)}$ term using standard analytic bootstrap methods. We then compute the one loop $G^{(R|F^2)}$ term using the AdS unitarity-cut method \cite{Aharony:2016dwx}, which fixes the one loop amplitude in terms of the tree level anomalous dimensions of double trace operators. These double trace operators are degenerate at tree level, and unmixing them requires us to compute tree level correlators of the form $\langle\bft 2 pp\rangle$ and $\langle\bft \bft pp\rangle$, where $\bft$ denotes the lowest graviton mode and $p$ denotes the $p$-th lowest gluon KK mode. We then take the flat space limit \cite{Penedones:2010ue} of $G^{(R|F^2)}$ and match it to the corresponding flat space amplitude \cite{Porkert:2022efy}.

To fix the contact terms $G^{(RF^2)}$, $G^{(R^2F^2)}$, and $G^{(\log)}$, we make use of the general relation in \cite{Chester:2022sqb} between an integral of 4d $\mathcal{N}=2$ flavor multiplet correlators and derivatives of the mass deformed $S^4$ free energy $F$. In our case we have the mass derivatives $\partial_\mu^2\partial_m^2F(\mu,m)\vert_{\mu=m=0}$, where $\mu$ and $m$ are masses for the $SO(8)$ and $SU(2)_L$ flavor symmetries, respectively. This quantity was computed using supersymmetric localization in terms of an $2N$-dimensional matrix model integral \cite{Pestun:2016zxk}, which is a modular invariant function of $\tau$.\footnote{If one takes derivatives of the four different $SO(8)$ masses $\mu_i$ as considered in \cite{Behan:2023fqq}, then the observable is now covariant under $SL(2,\mathbb{Z})$.} We compute $\partial_\mu^2\partial_m^2F(\mu,m)\vert_{\mu=m=0}$ at large $N$ and finite $\tau$, and give evidence that it takes the form
\es{Fprimetaufinal}{
&-\partial_\mu^2\partial_m^2F(\mu,m)\vert_{\mu=m=0}=16N-12\log N-12E(1,\tau)+c_{\mathrm{F}_2}-\sqrt{\frac{2}{N}}E(3/2,\tau)\\
&\qquad\qquad-\frac{31}{6N}+
\frac{1}{2\sqrt{2}N^{3/2}}\left[E(3/2,\tau)-\frac{3}{8}E(5/2,\tau)\right]+\frac{c_{\mathrm{F}_4}}{N^2}\\
&\qquad\qquad+\frac{9}{64\sqrt{2}N^{5/2}}
\left[-\frac{45}{64}E(3/2,\tau)+E(5/2,\tau)-\frac{45}{64}E(7/2,\tau)\right]+O(1/N^{7/2})\,,
}
where we did not analytically determine the $\tau$-independent constants $c_{\mathrm{F}_2}$ and $c_{\mathrm{F}_4}$, and $E(r,\tau)$ are weight $r$ non-holomorphic Eisenstein series.\footnote{For $r=1$, we perform a certain regularization as described in the main text that preserves modular invariance.} This observable takes a similar form as the integrated correlator of two gravitons and two giant graviton D3 branes in $AdS_5\times S^5$ \cite{Brown:2024tru}. We use this constraint to show that $G^{(RF^2)}$ vanishes. We then combine this constraint with the known flat space coefficient $G^{(R^2F^2)}$ \cite{Kiritsis:2000zi}, which is proportional to $E(1,\tau)$, to fix the $\tau$-dependence of $G^{(R^2F^2)}$. Note that $G^{(R^2F^2)}$ does not vanish, unlike the M-theory case considered in  \cite{Chester:2023qwo}. We similarly fix the $\tau$-independent logarithmic threshold $G^{(\log)}$. 

The rest of this paper is organized as follows. In Section \ref{sec:setup} we discuss constraints from superconformal symmetry on the various correlators we consider, as well as their large $N$ expansion and integrated constraints. In Section \ref{sec:correlators} we compute the various holographic correlators in the large $N$ expansion, including the one loop correction $G^{(R|F^2)}$. In Section \ref{sec:loc} we compute $\partial_\mu^2\partial_m^2F(\mu,m)\vert_{\mu=m=0}$ at large $N$ and finite $\tau$. In Section \ref{sec:glugrascattering} we use this localization constraint as well as the flat space limit to fix $G^{(RF^2)}$ and $G^{(R^2F^2)}$. We finish with a discussion of future directions in Section \ref{sec:conclusion}. Technical details of the calculations are given in the various Appendices, and we include an attached \texttt{Mathematica} notebook with further details.

\section{Setup}\label{sec:setup}

In this work we consider a 4d $\mathcal{N}=2$ SCFT with $SO(8)\times SU(2)_L$ global symmetry, which admits a description as an $USp(2N)$ gauge theory with 8 fundamental half-hypermultiplets and two half-hypermultiplets in the irreducible antisymmetric representation of $USp(2N)$ -- see, {\it e.g.}, \cite{Bedford:2007qj} for a detailed description of the field content. This setup can be realized in F-theory by placing $N$ D3 branes probing a sevenbrane $D_4$ singularity in F-theory \cite{Sen:1996vd,Banks:1996nj}, which leads to the holographic description of \cite{Fayyazuddin:1998fb,Aharony:1998xz}, where the geometry near the D3 branes is AdS$_5\times S^5/\mathbb{Z}_2$ and the $\mathbb{Z}_2$ action has an AdS$_5\times S^3$ fixed point along which the sevenbranes are extended. The main observable that we consider in this work is the mixed four-point function of two $SO(8)$ and two $SU(2)_L$ moment map operators, which is holographically dual to a mixed scattering amplitude between two gluons and two gravitons. In this section we begin by introducing the constraints of superconformal symmetry on this and other four-point functions that will be useful in this paper. We then discuss the large $N$ expansion of this observable, discussing which parts we will compute and laying down a plan for the rest of the paper. Finally, we present the integrated constraint introduced in \cite{Chester:2022sqb} for 4d $\mathcal{N}=2$ theories,  that we will use to determine certain coefficients appearing in the expression of the holographic correlator that cannot be fixed by bootstrap considerations.

\subsection{Superconformal kinematics}

Let us begin by establishing some conventions. We will use capitalized latin indices $A=1,2,\ldots,28$ for the adjoint representation of $SO(8)$, plain greek indices $\alpha,\beta=1,2$ for the fundamental representation of the $SU(2)_R$ factor of the R-symmetry group (the rest being $U(1)_r$) and barred greek indices $\bar\alpha,\bar \beta=1,2$ for the fundamental representation of the $SU(2)_L$ factor of the global symmetry. The various central charges in this theory are, in the conventions of \cite{Aharony:2007dj}
\es{}{
k_{SO(8)}=4N\,,\quad
k_{SU(2)_L}=2N^2-N-1\,,\quad
a=\frac{N^2}{2}+\frac{N}{2}-\frac{1}{24}\,,\quad
c=\frac{N^2}{2}+\frac{3N}{4}-\frac{1}{12}\,,
}
where $k_G$ is the flavor central charge associated with the factor $G$ of the global symmetry, while $a$ and $c$ are the usual stress-tensor central charges.

While the main object of our investigation is a moment map four-point function, we are going to consider four-point functions between the superconformal primaries of more general half-BPS supermultiplets\footnote{${B}_1{\bar{B}}_1$ multiplets in the notation of \cite{Cordova:2016emh}, $\hat{\mathcal{B}}_{p/2}$ multiplets in the notation of \cite{Dolan:2002zh}.}, which come in families labelled by an integer $p=2,3,\ldots$, associated with the representation of the superprimary under $SU(2)_R$. The case $p=2$ corresponds to the flavor current multiplet, whose superconformal primary is the moment map. Such superprimaries are scalars of the Lorentz group, and since their $U(1)_r$ charge vanishes only operators with vanishing $U(1)_r$ charge can be exchanged in their correlation functions. We summarize in Table \ref{tab:halfBPSmult} the operator content of half-BPS supermultiplets, only accounting for operators with zero $U(1)_r$ charge. 
\begin{table}[h!]
\centering
\begin{tabular}{|c  ||c |c| c|} 
\hline
component field & $s_p$ & $(A_p)_{\mu}$ & $r_p$\\
 \hline\hline
Lorentz spin $\ell$ & 0 & 1 & 0\\
\hline
conformal dimension $\Delta$ & $p$ & $p+1$ & $p+2$\\
\hline
$SU(2)_R$ spin $j_R$ & $\tfrac{p}{2}$ &$\tfrac{p}{2}-1$ & $\tfrac{p}{2}-2$\\
 \hline
\end{tabular}
\caption{Operators in half-BPS multiplets with vanishing $U(1)_r$ charge: for each $p$, $s_p$ and $r_p$ are Lorentz scalars, while $A_{\mu}$ is a vector. }
\label{tab:halfBPSmult}
\end{table}
Another relevant supermultiplet is the stress tensor multiplet $\mathcal{T}$\footnote{$A_2\bar{A}_{\bar{2}}$ in the notation of \cite{Cordova:2016emh}, $\hat{\mathcal{C}}_{0(0,0)}$ in the notation of \cite{Dolan:2002zh}.}: the operators with vanishing $U(1)_r$ charge in this multiplet are summarized in Table \ref{tab:stresstensormult}, while they are all singlets under the $SO(8)\times SU(2)_L$ global symmetry.
\begin{table}[h!]
\centering
\begin{tabular}{|c  ||c |c| c|} 
\hline
component field & $t$ & $B_{\mu}$ & $T_{\mu\nu}$\\
 \hline\hline
Lorentz spin $\ell$ & 0 & 1 & 2\\
\hline
conformal dimension $\Delta$ & 2 & 3 & 4\\
\hline
$SU(2)_R$ spin $j_R$ & 0 & 1 & 0\\
 \hline
\end{tabular}
\caption{Operators in the stress tensor multiplet with vanishing $U(1)_r$ charge.  All operators are singlets under $SO(8)\times SU(2)_L$.}
\label{tab:stresstensormult}
\end{table}

We are going to consider mixed correlators between superprimaries of supermultiplets of the first type,  that can be distinguished in two groups.
\paragraph{Gluons.} In the holographic description of the theory, one has an 8d SYM theory with $SO(8)$ gauge group on the worldvolume of the sevenbranes. The corresponding AdS$_5$ fields are obtained from their Kaluza-Klein (KK) reduction on $S^3$, which gives rise to a tower of KK modes labelled by an integer $p=2,3,\ldots$. For each $p$, these fields are organized in half-BPS supermultiplets of the type described in Table \ref{tab:halfBPSmult}, where in this case each field transforms in the adjoint representation of $SO(8)$ and in the spin $j_L=\tfrac{p}{2}-1$ representation of $SU(2)_L$ \cite{Aharony:1998xz}. We shall refer to these as super gluon multiplets. The case $p=2$, which is a singlet of $SU(2)_L$, corresponds to a massless $SO(8)$ vector supermultiplet in AdS$_5$, which is dual to the $SO(8)$ flavor current multiplet in the CFT. Using $SU(2)_R$ spinor polarizations $y^{\alpha}$ and $SU(2)_L$ spinor polarizations $\by^{\balpha}$, the superconformal primaries of these multiplets can be described by functions of the spacetime position $x$ and polarizations $y,\by$ via
\es{}{
\co^A_p(x;y,\by)=
\co^{A;\alpha_1,\ldots,\alpha_p; \balpha_1,\ldots,\balpha_{p-2}}(x)
y^{\beta_1}\ldots y^{\beta_p} \epsilon_{\alpha_1\beta_1}\ldots\epsilon_{\alpha_p\beta_p} 
\by^{\bbeta_1}\ldots \by^{\bbeta_{p-2}} \epsilon_{\balpha_1\bbeta_1}\ldots\epsilon_{\balpha_{p-2}\bbeta_{p-2}}\,, 
}
where $\epsilon$ is the two-dimensional antisymmetric tensor.

\paragraph{Gravitons.} The KK reduction of the ten-dimensional graviton to AdS$_5$ gives rise to a multitude of fields with Lorentz spin ranging from zero to two. A subset of these modes can also be organized into half-BPS supermultiplets of the type described in Table \ref{tab:halfBPSmult}, although sitting in different representations of the global symmetry compared to the gluons discussed above. To distinguish them from super gluons, we shall use boldface letters for super graviton multiplets, which are also labelled by an integer $\bfp=2,3,\ldots$ and are singlets of $SO(8)$, while having $SU(2)_L$ spin $j_L=\tfrac{\bfp}{2}$ (same as the $SU(2)_R$ spin $j_R$ of the superprimary in the multiplet). The case $\bfp=2$ corresponds to a 5d massless vector multiplet, dual to the $SU(2)_L$ flavor current supermultiplet. Note that we still refer to these as super graviton multiplets, although there are no spin-two fields, to remind of their origin from the ten-dimensional graviton. On the other hand, the massless 5d graviton is dual to the stress tensor, which sits in the multiplet described in Table \ref{tab:stresstensormult}. Introducing polarizations as above, we describe the superprimary of these super graviton multiplets as
\es{}{
\bfo_{\bfp}(x;y,\by)=
\bfo^{\alpha_1,\ldots,\alpha_{\bfp}; \balpha_1,\ldots,\balpha_{\bfp}}(x)
y^{\beta_1}\ldots y^{\beta_{\bfp}} \epsilon_{\alpha_1\beta_1}\ldots\epsilon_{\alpha_{\bfp}\beta_{\bfp}} 
\by^{\bbeta_1}\ldots \by^{\bbeta_{\bfp}} \epsilon_{\balpha_1\bbeta_1}\ldots\epsilon_{\balpha_{\bfp}\bbeta_{\bfp}}\,.
}

The main purpose of this work is to study the large $N$ expansion of the four-point function between two $SU(2)_L$ and two $SO(8)$ moment map operators,
\es{GGgg_2222}{
\langle \bft\bft 22 \rangle^{AB} \equiv 
\langle \bfo_{\bft}(1)\bfo_{\bft}(2)\co^A_2(3)\co^B_2(4)\rangle =\frac{(y_1\cdot y_2)^2(y_3\cdot y_4)^{2}}{x_{12}^4x_{34}^{4}}(\by_1\cdot \by_2)^{2}\delta^{AB}\,G(U,V;\alpha)\,,
}
where $G$ is a degree-two polynomial in $\alpha$ and we have introduced the cross ratios
\es{crdef}{
U=\frac{x_{12}^2 x_{34}^2}{x_{13}^2 x_{24}^2}=z \bar{z}\,, \quad V=\frac{x_{14}^2 x_{23}^2}{x_{13}^2 x_{24}^2}=(1-z)(1-\bar{z})\,,\quad
\alpha=\frac{\left(y_1 \cdot y_3\right)\left(y_2 \cdot y_4\right)}{\left(y_1 \cdot y_2\right)\left(y_3 \cdot y_4\right)}\,,
}
and $y_i \cdot y_j=\epsilon_{\alpha\beta}y_i^\alpha y_j^\beta$. Superconformal invariance implies that the function $G$ appearing in \eqref{GGgg_2222} should satisfy the superconformal Ward identities \cite{Dolan:2001tt}, which can be formally solved writing
\es{G_SCWI}{
G(U,V;\alpha)=\frac{z(1-\alpha \bar{z})f(\bar{z})-\bar{z}(1-\alpha z)f({z})}{z-\bar{z}}+(1-\alpha z)(1-\alpha\bar{z})\mathcal{G}(U,V)\,,
}
where $U=z\bar{z}$, $V=(1-z)(1-\bar{z})$ and we refer to $\mathcal{G}$ as the reduced correlator and to $f$ as the holomorphic correlator, which is protected and can be computed from the chiral algebra \cite{Beem:2013sza}. Note that both functions are independent of $\alpha$.

Although this is the object that we ultimately want to compute, as well known one of the artifacts of the large $N$ expansion in holographic CFTs is the presence of operators mixing. To address this, we are going to consider two families of mixed four-point functions. The first, $\langle \bft \bft pp\rangle$, is a generalization of \eqref{GGgg_2222}, in that it reduces to the latter for $p=2$. In the general case, however, there are three distinct $SU(2)_L$ channels, corresponding to the exchange of operators with spin $j_L=0,1,2$ in the direct OPE channel. This is restricted to $j_L=0,1$ for $p=3$ and to $j_L=0$ for $p=2$. In the general case, we set
\es{GGgg_22pp}{
\langle \bft \bft p p \rangle^{AB} &\equiv 
\langle \bfo_{\bft}(1)\bfo_{\bft}(2)\co^A_{p}(3)\co^B_{p}(4)\rangle
\\
&=\frac{(y_1\cdot y_2)^2(y_3\cdot y_4)^{p}}{x_{12}^4x_{34}^{2p}}\delta^{AB}\bar{y}_{34}^{p-4} G_{\bft \bft p p}(U,V;\alpha,\bar{y})\,,\\
G_{\bft \bft p p}(U,V;\alpha,\bar{y})&=\left[
\bar{y}_{12}^2\bar{y}_{34}^2
G^{(12)}_{\bft \bft p p}(U,V;\alpha) +
\bar{y}_{13}^2\bar{y}_{24}^2
G^{(13)}_{\bft \bft p p}(U,V;\alpha)+
\bar{y}_{14}^2\bar{y}_{23}^2
G^{(14)}_{\bft \bft p p}(U,V;\alpha)\right]\,,
}
where while for $p>3$ the three functions $G^{(12)}$, $G^{(13)}$ and $G^{(14)}$ are independent, for the special cases $p=2,3$ we have\footnote{Note that for $p=3$ with the restriction $G^{(14)}=-G^{(13)}$ we can rewrite \eqref{GGgg_22pp} as
\es{}{
\langle \bft \bft p p \rangle^{AB} &=\delta^{AB}\mathcal{K}_{22pp}\bar{y}_{12}\left[
\bar{y}_{12}\bar{y}_{34}
G^{(12)}_{\bft \bft p p}(U,V;\alpha) +
(\bar{y}_{13}\bar{y}_{24}-\bar{y}_{14}\bar{y}_{23})
G^{(13)}_{\bft \bft p p}(U,V;\alpha)\right]\,,
}
which manifestly has the correct degree in all the $\bar{y}_i$ coordinates.}
\es{}{
p=2&:\quad G^{(14)}_{\bft \bft p p}=G^{(13)}_{\bft \bft p p}=0\,,\\
p=3&:\quad G^{(14)}_{\bft \bft p p}=-G^{(13)}_{\bft \bft p p}\,.
}
It is also sometimes convenient to write \eqref{GGgg_22pp} in terms of the three exchanged irreducible representations of $SU(2)_L$ -- the singlet (S),    antisymmetric (A) and symmetric traceless (T) of $SO(3)\simeq SU(2)/\Z_2$:
\es{GGgg_22pp_reps}{
G_{\bft \bft p p}(U,V;\alpha,\bar{y})=\sum_{I\in \{S,T,A\}}Q_{(I)}G^{(I)}_{\bft \bft p p}\,,
}
where
\es{}{
Q_{(S)}&=\bar{y}_{12}^2\bar{y}_{34}^2\,,\\
Q_{(A)}&=\frac{1}{2}\left[\bar{y}_{14}^2\bar{y}_{23}^2-\bar{y}_{13}^2\bar{y}_{24}^2\right]\,,\\
Q_{(T)}&=\frac{1}{2}\left[\bar{y}_{14}^2\bar{y}_{23}^2+\bar{y}_{13}^2\bar{y}_{24}^2\right]-\frac{1}{3}\bar{y}_{12}^2\bar{y}_{34}^2\,,
}
and note that the relation between the two sets of functions we just introduced is
\es{}{
G^{(S)}_{\bft \bft p p}=\frac{1}{3}(G^{(13)}_{\bft \bft p p}+G^{(14)}_{\bft \bft p p})+G^{(12)}_{\bft \bft p p}\,,\quad
G^{(A)}_{\bft \bft p p}=G^{(14)}_{\bft \bft p p}-G^{(13)}_{\bft \bft p p}\,,\quad
G^{(T)}_{\bft \bft p p}=G^{(14)}_{\bft \bft p p}+G^{(13)}_{\bft \bft p p}\,.
}
Note that for $p=2$ we only have the singlet $G^{(S)}$ while for $p=3$ the symmetric traceless contribution $G^{(T)}=0$, as expected from the fact that in the $s$-channel OPE only $j_L=0,1$ are exchanged.

Finally, the last correlator of interest is $\langle \bft 2 pp\rangle$, which vanishes for $p=2$ while for $p>2$ we can make a singlet by taking the $j_L=1$ component in the $p\times p$ OPE. Thus we only have one independent function and we define
\es{Gggg_22pp}{
\langle  \bft 2 p p \rangle^{ABC} \equiv 
\langle  \bfo_{\bft}(1)\co_{2}(2)^A\co^B_{p}(3)\co^C_{p}(4)\rangle=f^{ABC}\frac{(y_1\cdot y_2)^2(y_3\cdot y_4)^{p}}{x_{12}^4x_{34}^{2p}}\bar{y}_{13}\bar{y}_{14}\bar{y}_{34}^{p-3}G_{\bft 2 p p}(U,V;\alpha)\,.
}
To conclude, we observe that the superconformal Ward identities should be satisfied individually by each of the four functions -- $G^{(12)}_{\bft \bft p p}, G^{(13)}_{\bft \bft p p}, G^{(14)}_{\bft \bft p p}, G_{\bft 2 p p}$ -- introduced to describe these last two classes of correlators. As a result, for each of those functions $G$ there is a corresponding $(f,\mathcal{G})$ pair related to $G$ as in \eqref{G_SCWI}.

\subsection{Large $N$ expansion}

The main purpose of our work is to study the mixed gluon-graviton four-point function \eqref{GGgg_2222} in a large $N$ expansion. From the point of view of string theory, the system is described by the action
\es{S_total}{
S&=S_{\text{SUGRA}}[G_{MN},\ldots]+S_{\text{SYM}}[A_{\mu},\ldots;\,g_{\mu\nu},\ldots]\,,\\
S_{\text{SUGRA}}[G_{MN},\ldots]&=\frac{1}{2\kappa^2_{\mathrm{10d}}}\int d^{10}X\,\sqrt{-G}\,[R+\mathrm{HD}+\mathrm{SUSY}]\,,\\
S_{\text{SYM}}[A_{\mu},\ldots;\,g_{\mu\nu},\ldots]&=-\frac{1}{4\mathtt{g}^2_{\text{8d}}}\int d^8 x \sqrt{-g}\,\left[F^A_{\mu\nu}F^{A,\mu\nu}+\mathrm{HD}+\mathrm{SUSY}\right]\,,
}
where $G_{MN}$ is the ten-dimensional metric and $g_{\mu\nu}$ is its pull-back on the worldvolume of the sevenbranes, while $F^A_{\mu\nu}$ denotes the field strength for the $SO(8)$ gauge theory. Here ``HD'' denotes higher-derivative terms in the low energy expansion of the effective action and ``SUSY'' denotes the supersymmetric completion of the given terms. In \eqref{S_total}, we have introduced the ten-dimensional gravitational coupling constant $\kappa_{\mathrm{10d}}$ and the eight-dimensional YM coupling constant $\mathtt{g}_{\text{8d}}$, which are related to the string parameters by
\es{kappagYM}{
2\kappa^2_{\mathrm{10d}}=(2\pi)^7g_s^2\ell_s^8\,,\quad
\mathtt{g}^2_{\text{8d}}=\frac{4\pi^4}{N}=(2\pi)^5g_s\ell_s^4\,,
}
where $g_s$ is the string coupling and $\ell_s$ the string length. Moreover, flux quantization relates the AdS radius $L$ to the number $N$ of D3 branes as in \eqref{adscft}. The large $N$ expansion of \eqref{GGgg_2222} then takes the form shown in \eqref{introG}.
%\es{introG}{
%G=G^{(0)}+\frac{1}{N^2}G^{(R)}+\frac{1}{N^3}\left[G^{(R|F^2)}+G^{(R^2F^2)}(\tau)+G^{(\log)}\,\log N\right]+O(1/N^4)\,,
%}
%where $G^{(R)}$ denotes the tree-level exchange of a graviton and $G^{(R|F^2)}$ is a one-loop contribution where the fields running inside the loop are gluons. On the other hand, $G^{(R^2F^2)}(\tau)$ is a six derivative contact term and $G^{(\log)}$ is another contact term whose role is to regularize, via its $\log N$ coefficient, the logarithmic divergence of the one-loop amplitude. We also note that, in principle, symmetry allows for the presence of a contact term of the schematic type $RF^2$ at $O(1/N^{5/2})$. However, this is not present in the flat space limit (see \cite{Stieberger:2009hq}) and we will prove that it is absent in AdS as well using supersymmetric localization. This is analogous to what happens for the M-theory setup considered in \cite{Chester:2023qwo}.

Our goal in the rest of the paper is to compute each term in \eqref{introG}. The graviton exchange term $G^{(R)}$ is rather simple to compute and can be normalized using protected OPE coefficients. The one loop term is the technically most challenging one, as it requires solving a mixing problem, but it can be computed using standard techniques \cite{Aharony:2016dwx} up to an additive renormalization ambiguity. On the other hand, $G^{(R^2F^2)}(\tau)$ and $G^{(\log)}$ are simple polynomials in Mellin space (which we introduce below), as they correspond to contact terms, but the coefficients in these polynomials cannot be computed using bootstrap considerations alone. We have emphasized that $G^{(R^2F^2)}(\tau)$ is a function of the axio-dilaton field
\es{}{
\tau=\frac{\chi}{2\pi}+\frac{i}{g_s}\,,
}
where $\chi$ is the axion. 

As well known, the computation of holographic correlators is greatly simplified by the use of Mellin space. Moreover, to solve the one loop mixing problem we are going to consider the correlators $\langle \bft \bft pp\rangle$ and $\langle \bft \bft pp\rangle$, so it will be convenient to introduce a Mellin representation for generic correlation functions between four scalars of dimensions $(\Delta_1,\Delta_2,\Delta_3,\Delta_4)=(2,2,p,p)$, which can be then adapted to our main correlator \eqref{GGgg_2222} simply by setting $p=2$. For any of the three functions $G$ appearing in \eqref{GGgg_22pp} or for the function $G$ in \eqref{Gggg_22pp}, we set
\es{}{
G(U,V;\alpha)&=\int_{-i\,\infty}^{+i\,\infty}\frac{ds dt}{(4\pi i)^2}U^{\tfrac{s}{2}}V^{\tfrac{t-p-2}{2}}\mathbf{\Gamma}\, M(s,t;\alpha)\,,\\
\mathbf{\Gamma}&=\Gamma\left[2-\tfrac{s}{2}\right]\Gamma\left[p-\tfrac{s}{2}\right]\Gamma\left[\tfrac{2+p-t}{2}\right]^2\Gamma\left[\tfrac{2+p-u}{2}\right]^2\,,
}
where $s+t+u=2(2+p)$. On the other hand, for the reduced correlator $\mathcal{G}$ associated to each of those functions, we set
\es{}{
\mathcal{G}(U,V)&=\int_{-i\,\infty}^{+i\,\infty}\frac{ds dt}{(4\pi i)^2}U^{\tfrac{s}{2}}V^{\tfrac{t-p-2}{2}}\mathbf{\widetilde{\Gamma}}\, \mathcal{M}(s,t)\,,\\
\mathbf{\widetilde{\Gamma}}&=\Gamma\left[2-\tfrac{s}{2}\right]\Gamma\left[p-\tfrac{s}{2}\right]\Gamma\left[\tfrac{2+p-t}{2}\right]^2\Gamma\left[\tfrac{2+p-\tilde{u}}{2}\right]^2\,,
}
where $s+t+\tilde{u}=2(1+p)$. The integration contour in both cases includes all poles in $s$ and $t$, but not those in $u$ or $\tilde{u}$. 

To fix the coefficients appearing in $G^{(R^2F^2)}(\tau)$ and $G^{(\log)}$ we are going to use two additional ingredients. One is the flat space limit \cite{Penedones:2010ue}, which relates the reduced Mellin amplitude $\mathcal{M}$ corresponding to the correlator \eqref{GGgg_2222} to a flat space scattering amplitude between two gravitons and two gluons via 
\es{flatspacelim}{
\mathcal{A}(s,t)=\lim_{L\to\infty}  -\pi^5L^6\int\frac{d\beta}{2\pi i}\frac{e^\beta}{\beta^4}\frac{L^4(u+\alpha s)^2}{32}\mathcal{M}\Big(\frac{L^2}{2\beta}s,\frac{L^2}{2\beta}t\Big)\,,
}
where $\mathcal{A}(s,t)$ is the {\it ten}-dimensional scattering amplitude in flat space. Note that in the flat space limit we should think of this as a scattering between gluons confined on the eight-dimensional sevenbranes and gravitons propagating in the whole ten-dimensional spacetime. This is why we have chosen the power of $L$ in \eqref{flatspacelim} to be that for a ten-dimensional scattering process, as opposed to the case of pure gluon scattering considered in \cite{Alday:2021odx, Alday:2021ajh,Behan:2023fqq}, which is genuinely eight-dimensional. See \eqref{A_flat_fromM} and below for more comments on the interplay between eight and ten dimensions in the explicit results, specifically for one loop diagrams.

The other ingredient is the integrated correlator constraint introduced in \cite{Chester:2022sqb}, which we review in the next subsection.

\subsection{Integrated constraint}

Using supersymmetric localization \cite{Pestun:2007rz} it is possible to compute the mass deformed sphere free energy $F(\mu_i,m)$ of the theory, where $\mu_i$, $i=1,\ldots,4$ are four masses for the fundamental hypermultiplets along the Cartan generators of $SO(8)$ and $m$ is a mass for the antisymmetric hypermultiplet, along the Cartan of $SU(2)_L$. The case of four independent $\mu_i$ with $m=0$ was considered in \cite{Behan:2023fqq}, while here we are interested in the case of four identical $SO(8)$ masses $\mu_i\equiv \mu$ and $m\neq 0$. The constraint derived in \cite{Chester:2022sqb} relates mass derivatives of $F(\mu,m)$ to a certain integral of the connected part of the correlator \eqref{GGgg_2222}. More precisely, in our case the constraint reads
\es{intconstraint}{
-\left. \partial^2_\mu\partial^2_m F\right|_{\mu=m=0}=8k_{SU(2)_L}k_{SO(8)}I[\mathcal{G}]\,,
}
where the integral is defined as
\es{intG}{
 I[\mathcal{G}]\equiv\frac{1}{ \pi}  \int dR\, d\theta\, R^3 \sin^2 \theta
     \frac{\bar D_{1,1,1,1}(U,V)\cG(U, V)}{U^2} 
\bigg|_{\substack{U = 1 + R^2 - 2 R \cos \theta \\
    V = R^2}}\,,
    }
and the function $\bar D_{1,1,1,1}(U,V)$ is given by
\es{db1111}{
\bar{D}_{1,1,1,1}(U,V) = \frac{1}{z - \zb} \left ( \log(z\zb) \log \frac{1 - z}{1 - \zb} + 2\text{Li}(z) - 2\text{Li}(\zb) \right )\,.
}
It is also convenient to express \eqref{intG} as an integral of the reduced Mellin amplitude $\mathcal{M}$ associated with the reduced correlator $\mathcal{G}$. The result, found in \cite{future}, is
\es{intM}{
  I[\mathcal{M}]\equiv -\int \frac{ds dt}{(4\pi i)^2}& \Bigg[\mathcal{M}(s,t) \Gamma[2-s/2]\Gamma[s/2]\Gamma[2-t/2]\Gamma[t/2]\Gamma[2-{u}/2]\Gamma[{u}/2] \\
 & \times \Big(\frac{H_{\frac s2-1}+H_{1-\frac s2}}{(t-2)(u-2)}+\frac{H_{\frac t2-1}+H_{1-\frac t2}}{(s-2)(u-2)}+\frac{H_{\frac u2-1}+H_{1-\frac u2}}{(s-2)(t-2)}\Big)\Bigg]\,,
}
where $H_n$ is a harmonic number.

\section{Holographic correlators}\label{sec:correlators}

The goal of this section is to present the various ingredients appearing at the first few orders in the large $N$ expansion of the $\langle \bft \bft 22\rangle$ correlator, see \eqref{introG}. The technically most challenging part is given by the computation of the one loop term $G^{(R|F^2)}$, since it requires solving a mixing problem which in turn necessitates the computation of additional correlators at tree level: $\langle \bft \bft pp\rangle$ and $\langle \bft 2 pp\rangle$. We will therefore start with that computation and then proceed to discuss the mixing problem and the derivation of the one loop term. We will conclude with a brief presentation of the higher-derivative corrections we are interested in, which introduce the free parameters that we will fix in the rest of the paper.

\subsection{Tree level}

As explained, we are interested in computing two types of correlators at tree level. In both cases, we will consider the exchange of the three types of multiplets introduced in Section \ref{sec:setup}: the stress tensor multiplet $\mathcal{T}$, super gluon multiplets $\co_p$ and super graviton multiplets $\bfo_\bfp$. 

\subsubsection{Two gravitons, two gluons: $\langle\bft \bft pp\rangle$.} Let us consider the two independent OPE channels for $\langle \bft \bft p p\rangle$, focusing on the single trace operators that can be exchanged since those correspond to exchange Witten diagrams.
\begin{itemize}
\item {\it Direct channel.} The $\bfo_\bft \times \bfo_\bft$ OPE contains $SO(8)$ singlets with $j_L\in\{0,1,2\}$ and $j_R\in\{0,1,2\}$. Among the three types of short multiplets listed above, this allows for the exchange of the stress tensor multiplet $\mathcal{T}$, as well as super graviton multiplets $\bfo_\bft$ and $\bfo_{\mathbf{4}}$. The latter, however, is an extremal coupling and therefore the corresponding Witten exchange diagram is absent. On the other hand, the $\co_p\times \co_p$ OPE contains both singlets and adjoints of $SO(8)$, as well as all $SU(2)_L$ spins $j_L\in\{0,\ldots,p-2\}$ and $SU(2)_R$ spins $j_R\in\{0,\ldots,p\}$. As a result, for $p=2$ only the stress tensor multiplet $\mathcal{T}$ is exchanged, while for $p>2$ both $\mathcal{T}$ and $\bfo_\bft$ are exchanged. Since both the gluon-gluon-graviton and the graviton-graviton-graviton vertices have couplings of order $1/N$, both exchange diagrams appear at order $1/N^2$.

\item {\it Crossed channel.} The other independent channel has the $\bfo_\bft\times \co_p$ OPE repeated twice. This contains $SO(8)$ adjoint fields with $j_L=\tfrac{p}{2}-2,\tfrac{p}{2}-1,\tfrac{p}{2}$ and $j_R=\tfrac{p}{2}-1,\tfrac{p}{2},\tfrac{p}{2}+1$: precisely the quantum numbers of $\mathcal{O}_{p-2}$, $\mathcal{O}_{p}$ and $\mathcal{O}_{p+2}$. However, the first and the third are extremal, so only $\mathcal{O}_{p}$ can be exchanged. Moreover, note that this is only possible for $p>2$ since for $p=2$ the only allowed $SU(2)_L$ spin is $j_L=1$ but $\mathcal{O}_{p}$ has $j_L=0$. So we have a gluon-gluon-graviton vertex repeated twice, which gives an exchange at order $1/N^2$.

\end{itemize}

To summarize, we expect
\es{Mtree_GGgg}{
\mathcal{M}^{(\text{tree})}_{\bft\bft pp}(s,t;\bar{y})=
\bar{y}_{12}^2\bar{y}_{34}^2\,\mathcal{M}^{(\mathcal{T})}(s,t)+\mathtt{t}_s\,\mathcal{M}^{(s)}(s,t)+\mathtt{t}_t\,\mathcal{M}^{(t)}(s,t)+\mathtt{t}_u\,\mathcal{M}^{(u)}(s,t)\,,
}
where $\mathtt{t}_i$, $i=s,t,u$, are three polynomials in the polarization $\bar{y}$, while $\mathcal{M}^{(i)}$ are reduced Mellin amplitudes corresponding to the exchange of $\bfo_\bfp$ (for the $s$ channel) or $\co_p$ (for the $t$ and $u$ channels). We now proceed to the computation of the various ingredients and note that only the stress tensor can be exchanged for $p=2$, so we expect the last three terms in \eqref{Mtree_GGgg} to vanish in that case.

The amplitude corresponding to the stress tensor is the simplest. It can be derived making an ansatz for the full Mellin amplitude as a sum of exchange diagrams corresponding to the three operators listed in Table \ref{tab:stresstensormult} and then demanding that it satisfies the superconformal Ward identities. Alternatively, one can just make an ansatz for the reduced correlator and then check that it gives the correct poles. Either way, it is easy enough to determine that
\es{}{
\mathcal{M}^{(\mathcal{T})}=\mathcal{N}_{\bft \bft pp}^{(\mathcal{T})}\frac{1}{s-2}\,,
}
where $\mathcal{N}_{\bft \bft pp}^{(\mathcal{T})}$ is a normalization term that we will fix later in this section.

Next, let us focus on the $SU(2)_L$ tensor structures $\mathtt{t}_i$, $i=s,t,u$. To derive those, it is useful to consider the three-point functions (see Appendix \ref{app:free} for the derivation)
\es{}{
\langle \bfo_\bft(x_1;y_1,\bar{y}_1)\bfo_\bft(x_2;y_2,\bar{y}_2)\bfo_\bft(x_3;y_3,\bar{y}_3)\rangle&=\frac{2\sqrt{2}}{\sqrt{k_{SU(2)_L}}}(12)(13)(23)\bar{y}_{12}\bar{y}_{13}\bar{y}_{23}\\
&=\frac{2}{N}(12)(13)(23)\bar{y}_{12}\bar{y}_{13}\bar{y}_{23}+O(1/N^2)\,,
}
and
\es{}{
\langle \cO_p^A(x_1;y_1,\bar{y}_1)\cO_p^B(x_2;y_2,\bar{y}_2)\bfo_\bft(x_3;y_3,\bar{y}_3)\rangle&=\frac{p-2}{\sqrt{2k_{SU(2)_L}}}\delta^{AB}(12)^{p-1}(13)(23)\bar{y}_{12}^{p-3}\bar{y}_{13}\bar{y}_{23}\\
&=\frac{p-2}{2N}\delta^{AB}(12)^{p-1}(13)(23)\bar{y}_{12}^{p-3}\bar{y}_{13}\bar{y}_{23}+O(1/N^2)\,.
}
The tensor structures associated with the various exchanges can be found by ``opening the indices'' corresponding to exchanged fields in the product of two three-point functions and summing over the open indices. We find
\es{}{
&\sum_{\bar{\alpha},\bar{\beta}}\left[\frac{\partial^2}{\partial \bar{y}^{\bar{\alpha}}_E\partial \bar{y}^{\bar{\beta}}_E}\bar{y}_{12}\bar{y}_{1E}\bar{y}_{2E}\right]\left[\frac{\partial^2}{\partial \bar{y}^{\bar{\alpha}}_E\partial \bar{y}^{\bar{\beta}}_E}\bar{y}_{34}^{p-3}\bar{y}_{3E}\bar{y}_{4E}\right]\\
&\propto \bar{y}_{12}\bar{y}_{34}^{p-3}\left(\bar{y}_{13}\bar{y}_{24}+\bar{y}_{14}\bar{y}_{23}\right)\equiv \mathtt{t}_s\,,
}
and
\es{}{
&\sum_{\bar{\alpha}_1,\ldots,\bar{\alpha}_{p-2}}\left[\frac{\partial^{p-2}}{\partial \bar{y}^{\bar{\alpha}_1}_E\cdots \partial \bar{y}^{\bar{\alpha}_{p-2}}_E}\bar{y}_{14}\bar{y}_{1E}\bar{y}_{4E}^{p-3}\right]\left[\frac{\partial^{p-2}}{\partial \bar{y}^{\bar{\alpha}_1}_E\cdots \partial \bar{y}^{\bar{\alpha}_{p-2}}_E}\bar{y}_{23}\bar{y}_{2E}\bar{y}_{3E}^{p-3}\right]\\
&\propto \bar{y}_{14}\bar{y}_{23}\bar{y}_{34}^{p-4}\left(\bar{y}_{14}\bar{y}_{23}-(p-2)\bar{y}_{13}\bar{y}_{24}\right) \equiv \mathtt{t}_t\,.
}
Similarly,  exchanging 1 and 2,
\es{}{
\mathtt{t}_u\equiv \bar{y}_{13}\bar{y}_{24}\bar{y}_{34}^{p-4}\left(\bar{y}_{13}\bar{y}_{24}-(p-2)\bar{y}_{14}\bar{y}_{23}\right)\,.
}
Note that the three structures are not independent, actually they satisfy
\es{}{
\mathtt{t}_s=\mathtt{t}_u-\mathtt{t}_t\,.
}
Since there are only two independent tensor structures, and using the symmetry under the exchange of the first two operators (that is, under $t\leftrightarrow \tilde{u}$), we can actually rewrite our ansatz as
\es{}{
\mathcal{M}^{(\text{tree})}_{\bft\bft pp}(s,t;\bar{y})=
\bar{y}_{12}^2\bar{y}_{34}^2\,\mathcal{M}^{(\mathcal{T})}(s,t)+\mathtt{t}_t\,\mathcal{M}_{st}+\mathtt{t}_u\,\mathcal{M}_{su}\,,
}
and since the exchanged and external fields transform in the same representation of the superconformal algebra, it turns out that the two independent reduced Mellin amplitudes $\mathcal{M}_{st}$ and $\mathcal{M}_{su}$ are exactly the same, up to an overall coefficient, as those computed in \cite{Alday:2021odx} in the context of pure gluon scattering in AdS, since they should satisfy the same constraints. Thus, we can ultimately write
\es{Mtree_GGgg_finalN}{
\mathcal{M}^{(\text{tree})}_{\bft\bft pp}(s,t;\bar{y})=
\mathcal{N}_{\bft \bft pp}^{(\mathcal{T})}\frac{\bar{y}_{12}^2\bar{y}_{34}^2}{s-2}+
\mathcal{N}_{\bft \bft pp}^{(\mathcal{J})}\left[\frac{\mathtt{t}_t}{(s-2)(t-p)}+\frac{\mathtt{t}_u}{(s-2)(\tilde{u}-p)}\right]\,,
}
where we remind that $s+t+\tilde{u}=2(1+p)$ and $\mathcal{N}_{\bft \bft pp}^{(\mathcal{J})}$ is another normalization term that we shall soon fix.

Converting \eqref{Mtree_GGgg_finalN} to spacetime, we find the function
\es{Ttree}{
\mathcal{T}^{(\text{tree})}_{\bft\bft pp}(U,V)=&-\mathcal{N}^{(\mathcal{T})}_{\bft \bft pp}\bar{y}_{12}^2\bar{y}_{34}^{2}\frac{U^p}{2}\bar{D}_{p+1,p+1,2,2}\\
&+\mathcal{N}^{(\mathcal{J})}_{\bft \bft pp}\bar{y}_{14}\bar{y}_{23}(\bar{y}_{14}\bar{y}_{23}-(p-2)\bar{y}_{13}\bar{y}_{24})\frac{U}{4}\bar{D}_{1,2,p+1,p}\\
&+\mathcal{N}^{(\mathcal{J})}_{\bft \bft pp}\bar{y}_{13}\bar{y}_{24}(\bar{y}_{13}\bar{y}_{24}-(p-2)\bar{y}_{14}\bar{y}_{23})\frac{U}{4}\bar{D}_{1,2,p,p+1}\,,
}
and we write the corresponding full correlator in spacetime as
\es{}{
G^{(\text{tree})}_{\bft\bft pp}(U,V;\alpha)=G^{(\text{free})}_{\bft\bft pp}(U,V;\alpha)+(1-\alpha z)(1-\alpha\bar{z})\mathcal{T}^{(\text{tree})}_{\bft\bft pp}(U,V)\,,
}
where $G^{(\text{free})}_{\bft\bft pp}(U,V;\alpha)$ is the free theory correlator computed in \eqref{free_22pp_GGgg}. On the other hand, we can also write the same correlator as
\es{}{
G^{(\text{tree})}_{\bft\bft pp}(U,V;\alpha)=\frac{z(1-\alpha \bar{z})f^{(\text{tree})}_{\bft\bft pp}(\bar{z})-\bar{z}(1-\alpha z)f^{(\text{tree})}_{\bft\bft pp}({z})}{z-\bar{z}}+(1-\alpha z)(1-\alpha\bar{z})\mathcal{G}^{(\text{tree})}_{\bft\bft pp}(U,V)\,,
}
which is convenient because it simplifies the expression of the OPE, which reads
\es{blocks_exp}{
\mathcal{G}_r(U,V)&=U^{-1}\left[\sum_{\ell,\Delta\ge \ell+2}\lambda^{(\bft,p)}_{\Delta,\ell,r}G_{\Delta+2,\ell}+\sum_{\ell}\lambda^{(\bft,p)}_{\ell+5,\ell+1,r}G_{\ell+5,\ell+1}+\lambda^{(\bft,p)}_{4,0,r}G_{4,0}\right]\,,\\
f_r(z)&=\sum_{\ell}\lambda^{(\bft,p)}_{\ell+5,\ell+1,r}k_{2\ell+6}+\lambda^{(\bft,p)}_{4,0,r}k_4+\delta_{r,\mathbf{1}}-\delta_{r,\mathbf{adj}}\lambda^{(\bft,p)}_{\mathcal{J}}\,k_2-\delta_{r,\mathbf{1}}\lambda^{(\bft,p)}_{\mathcal{T}}\,k_4\,,
}
in terms of 4d conformal blocks $G_{\Delta,\ell}$ and 1d conformal blocks $k_h$, where the subscript $r$ runs over the three representations in \eqref{GGgg_22pp_reps}. Each coefficient $\lambda^{(\bft,p)}_{-}$ is given as a product of OPE coefficients as $\lambda^{(\bft)}_{-}\lambda^{(p)}_{-}$, and for the current ($\mathcal{J}$) and stress tensor ($\mathcal{T}$) these can be computed combining our free theory results from Appendix \ref{app:free} with the results of \cite{Chang:2017xmr}. We find
\es{}{
\lambda^{(\bft,p)}_{\mathcal{J}}=\frac{2(p-2)}{N^2}\,,\quad
\lambda^{(\bft,p)}_{\mathcal{T}}=\frac{p}{6N^2}\,.
}
While in general the sum over long multiplets begins at twist two ($\Delta=\ell+2$), at large $N$ there should be no twist two long multiplets. This requirement fixes 1) the expression of $\mathcal{G}$ in terms of $\mathcal{T}$ and 2) the normalization terms in \eqref{Ttree}. We find
\es{}{
\mathcal{G}^{(\text{tree})}_{\bft\bft pp}(U,V)=\mathcal{T}^{(\text{tree})}_{\bft\bft pp}(U,V)+\frac{(p-3)^2}{N^2}U(\bar{y}_{13}^2\bar{y}_{24}^2+\bar{y}_{14}^2\bar{y}_{23}^2/V^2)\,,
}
while the normalizations are 
\es{normsGGgg22pp}{
\mathcal{N}^{(\mathcal{T})}_{\bft \bft pp}=-\frac{2p}{(p-2)!N^2}\,,\quad
\mathcal{N}^{(\mathcal{J})}_{\bft \bft pp}=-\frac{4}{(p-3)!N^2}\,.
}

In the special case $p=2$, as expected we find $\mathcal{N}^{(\mathcal{J})}_{\bft \bft 22}=0$ so that the correlator is completely determined by the exchange of the stress tensor. The result in this case gives the graviton exchange term $G^{(R)}$ appearing in \eqref{introG}, whose associated reduced Mellin amplitude is then
\es{MR}{
\mathcal{M}^{(R)}=-\frac{4}{s-2}\,.
}

\subsubsection{One gravitons, three gluons: $\langle\bft 2 pp\rangle$.}

Similarly to the previous case, let us consider the exchanged single trace operators in the two independent OPE channels to determine the associated exchanged Witten diagrams.

\begin{itemize}
\item {\it Direct channel.} Here we are interested in the single-trace operators appearing in the overlap between the OPEs $\bfo_\bft \times \co_2$ and $\co_p\times \co_p$. The former selects $SO(8)$ adjoint operators, so the only viable candidates are super gluon multiplets. However, the first OPE also forces $j_L=1$ for the putative exchanged gluon (and $j_R\in \{0,1,2\}$), allowing in principle only for the exchange of $\co_4$. But this is extremal so it cannot be exchanged: as a result, there are no exchange Witten diagrams in the direct channel.

\item {\it Crossed channel.} Focusing on $\bfo_\bft \times \co_p$ first, we find again that we have to focus on super gluons since only $SO(8)$ adjoints are exchanged. Neglecting extremal couplings, we find that $\co_p$ can be exchanged in this OPE and moreover it also appears in the $\co_2 \times \co_p$ OPE. Thus we have Witten diagrams corresponding to the exchange of $\co_p$ both in the $t$ and $u$ channel. They arise from the product of a gluon-gluon-graviton vertex at $O(1/N)$ at a gluon-gluon-gluon vertex at $O(1/N^{1/2})$, so this correlator appears at $O(1/N^{3/2})$.

\end{itemize}

As in the previous case, we have an exchange diagram which is required to satisfy all the kinematical constraints of the gluon exchange diagrams computed in \cite{Alday:2021odx}. This time, though, there are no exchanges in the $s$-channel. The result is then
\es{Mtree_Gggg}{
\mathcal{M}^{(\text{tree})}_{\bft 2 pp}=\frac{\mathcal{N}_{\bft 2 pp}}{(t-p)(\tilde{u}-p)}\,,
}
where $\mathcal{N}_{\bft 2 pp}$ is a normalization that we now proceed to fix. This reduced correlator contains no twist-two terms, so we have to proceed in a different way compared to the previous subsection. We observe that computing the full Mellin amplitude associated to \eqref{Mtree_Gggg} and summing over poles in $s,t$ to obtain the corresponding spacetime expression, we obtain the result for the full correlator
\es{}{
G^{(\text{tree})}_{\bft 2 pp}(U,V;\alpha)={\mathcal{N}_{\bft 2 pp}}\left[\frac{(p-2)!}{4}\frac{U^2}{V}\alpha(\alpha-1)+(1-\alpha z)(1-\alpha\bar{z}) \frac{U^2}{4}\bar{D}_{1,3,p,p}\right]\,,
} 
from which it is straightforward to derive the associated chiral algebra correlator
\es{}{
f^{(\text{tree})}_{\bft 2 pp}(z)=\mathcal{N}_{\bft 2 pp}\frac{(p-2)!}{4}\frac{z^2}{1-z}\,.
}
We can compare this to the expected result for $f^{(\text{tree})}_{\bft 2 pp}(z)$, which can be computed in the free theory and found to be
\es{}{
f^{(\text{free})}_{\bft 2 pp}(z)=\frac{p-2}{\sqrt{2}N^{3/2}}\frac{z^2}{1-z}\,,
}
see Appendix \ref{app:free} for details. Comparing the two fixes for us
\es{MtreeGggg}{
\mathcal{N}_{\bft 2 pp}=\frac{2\sqrt{2}}{(p-3)!N^{3/2}}\quad
\Rightarrow \quad
\mathcal{M}^{(\text{tree})}_{\bft 2 pp}=\frac{1}{(p-3)!N^{3/2}}\frac{2\sqrt{2}}{(t-p)(\tilde{u}-p)}\,.
}

\subsection{Mixing problem}

We would now like to compute the one loop term $G^{(R|F^2)}$ in \eqref{introG}, using the AdS unitarity method of \cite{Aharony:2016dwx}. This involves computing terms in the one loop correlator that are proportional to $\log^2 U$ (or $\log^2V$) using CFT data at previous perturbative orders. In particular, naively one is supposed to square the tree level anomalous dimensions. However, this squaring is complicated by the problem of operators mixing, which we now discuss in the two OPE channels separately.

\subsubsection{Direct channel} 

In the direct channel, the long supermultiplets exchanged in $\langle \bft \bft 22\rangle$ are $SU(2)_L\times SU(2)_R\times SO(8)$ singlets. Focusing on double trace operators made out of super gluon and super graviton multiplets, we find that for given twist $\tau=4+2n$ and spin $\ell$, there are $(2n+1)$ distinct operators that can be formed, of the schematic form
\es{}{
[\cO_2\cO_2]_n\,,\quad \ldots\,, \quad
[\cO_{n+2}\cO_{n+2}]_n\,,\quad 
[\bfo_{\bft}\bfo_{\bft}]_n\,,\quad \ldots\,, \quad
[\bfo_{\mathbf{n+2}}\bfo_{\mathbf{n+2}}]_n\,,
}
where for instance $[\mathcal{O}_p\mathcal{O}_p]_n\sim :\mathcal{O}_p \Box^{n+2-p} \partial^\ell\mathcal{O}_p:$ for given spin $\ell$ and twist $\tau=4+2n$. We adopt this as a basis of operators to span the degeneracy space, with a indices $A,B=1,\ldots,2(n+1)$ running over the $2(n+1)$ operators in the space. We also note that the basis naturally splits into two subspaces of equal dimension $n+1$: that of gluon-gluon ($O_{gg}$) double trace operators and that of graviton-graviton ($O_{\bfg \bfg}$) ones. We use indices $a,b=1,\ldots,n+1$ for the former and an indices $i,j=1,\ldots,n+1$ for the latter.

Clearly, given the arbitrary choice of basis, the dilatation operator of the theory in this basis will not necessarily be diagonal. Thus, we should really consider the anomalous dimensions as a matrix $\Gamma_{AB}$, whose entries at each perturbative order can be extracted from the perturbative expansion of
\es{Gammadef}{
\langle O_AO_B\rangle-\langle O_AO_B\rangle^{(0)}\,,
}
where the superscript zero in this case denotes the result in the generalized free theory (GFT) at $N=\infty$. We also note that the two-point function $\langle O_AO_B\rangle$ can be computed considering a four-point function between the four underlying single trace operators and taking suitable derivatives and coincident limits. This allows us to derive a few basic facts.

First, we consider GFT two-point function and define a matrix of norms
\es{}{
\langle O_AO_B\rangle^{(0)}\equiv \mathtt{g}_{AB}=
\begin{pmatrix}
\mathtt{g}_{ab} & 0\\
0 & \mathtt{g}_{ij}
\end{pmatrix}\,,
}
where the diagonal structure is inherited from the fact that mixed correlators $\langle \bfg\bfg gg\rangle$ are given by the identity in the GFT at $N=\infty$. Moreover, the same applies to $\langle ppqq\rangle$ and $\langle \bfp \bfp \bfq\bfq\rangle$ correlators for $p\neq q$, which implies that the matrices $\mathtt{g}_{ab}$ and $\mathtt{g}_{ij}$ are diagonal. We can then normalize the operators in such a way that $\mathtt{g}_{AB}$ is the identity. 

Next, we consider three-point functions with two single trace and one double trace operator in the GFT. For generic super gluons $g$, we find\footnote{In this subsection we will use the symbol $C$ to denote OPE coefficients between two short and one long operator. Products between these OPE coefficients give the coefficients $\lambda^{(-,-)}_{\Delta,\ell,r}$ of long multiplets appearing in \eqref{blocks_exp}.}
\es{}{
C^{(0)}_{ggA}=\langle gg {O}_A\rangle^{(0)}=(C^{(0)}_{gga},C^{(0)}_{ggi})=(C^{(0)}_{gga},0)\,,
}
where again we have used the fact that the three-point functions can be computed from GFT four-point functions where the last two points are taken to be coincident. Similarly, we have
\es{}{
C^{(0)}_{\bfg \bfg A}=\langle \bfg \bfg {O}_A\rangle^{(0)}=(C^{(0)}_{\bfg \bfg a},C^{(0)}_{\bfg \bfg i})=(0,C^{(0)}_{\bfg \bfg i})\,.
}

Last, we address the structure of the anomalous dimensions matrix $\Gamma_{AB}$. We can think of it in terms of $(n+1)\times (n+1)$ blocks
\es{}{
\Gamma_{AB}=
\begin{pmatrix}
\Gamma_{ab} & \Gamma_{aj}\\
\Gamma_{ib} & \Gamma_{ij}
\end{pmatrix}\,,
}
where note that $\Gamma_{ai}= \Gamma_{ia}$ and the leading order contributions, {\it i.e.} the tree-level anomalous dimensions we are interested in for the one loop computation, can be obtained respectively from
\es{}{
\langle gggg\rangle^{(\text{tree})}|_{\log U}&\to C^{(0)}_{ggA}\Gamma_{AB}^{(\text{tree})}C^{(0)}_{ggB}=C^{(0)}_{gga}\Gamma_{ab}^{(\text{tree})}C^{(0)}_{ggb}\,,\\
\langle gg\bfg\bfg\rangle^{(\text{tree})}|_{\log U}&\to C^{(0)}_{ggA}\Gamma_{AB}^{(\text{tree})}C^{(0)}_{\bfg\bfg B}=C^{(0)}_{gga}\Gamma_{aj}^{(\text{tree})}C^{(0)}_{\bfg\bfg j}\,,\\
\langle \bfg\bfg gg\rangle^{(\text{tree})}|_{\log U}&\to C^{(0)}_{\bfg\bfg A}\Gamma_{AB}^{(\text{tree})}C^{(0)}_{ggB}=C^{(0)}_{\bfg\bfg i}\Gamma_{ib}^{(\text{tree})}C^{(0)}_{ggb}\,,\\
\langle \bfg\bfg\bfg\bfg\rangle^{(\text{tree})}|_{\log U}&\to C^{(0)}_{\bfg\bfg A}\Gamma_{AB}^{(\text{tree})}C^{(0)}_{\bfg\bfg B}=C^{(0)}_{\bfg\bfg i}\Gamma_{ij}^{(\text{tree})}C^{(0)}_{\bfg\bfg j}\,,
}
from which we also derive that, given the large $N$ scaling of the associated correlators,
\es{}{
\Gamma_{ab}^{(\text{tree})}\sim O(1/N)\,,\quad
\Gamma_{ai}^{(\text{tree})}\sim\Gamma_{ia}^{(\text{tree})}\sim\Gamma_{ij}^{(\text{tree})}\sim O(1/N^2)\,.
}
The coefficient of each conformal block for the term proportional to $\log^2U$ in the direct channel OPE can be obtained by multiplying the square of the anomalous dimensions matrix with suitable OPE coefficients on the left and on the right. Using the results obtained so far, we find that the object we want to compute is
\es{Glooplog2U}{
\left. \mathcal{G}^{(R|F^2)}(U,V)\right|_{\log^2 U}=\sum_{n,\ell}F^{(s)}_{n,\ell}U^{-1}G_{6+2n+\ell,\ell}(U,V)\,,
}
where 
\es{log2directPRE}{
F^{(s)}_{n,\ell}=\frac{1}{N^3}\left[C^{(0)}_{\bft\bft i}\Gamma_{ia}^{(\text{tree})}\Gamma_{ab}^{(\text{tree})}C^{(0)}_{22b}\right]_{n,\ell}\,.
}
Following \cite{Behan:2022uqr}, we massage this expression by introducing the matrix
\es{}{
M_{ra}=\frac{C^{(0)}_{rra}}{\sqrt{\sum_b (C^{(0)}_{rrb})^2}}\,,
}
which is $(n+1)\times(n+1)$ since $r=2,\ldots,n+2$ runs over the $n+1$ supergluons $\mathcal{O}_2,\ldots, \mathcal{O}_{n+2}$ while $a=1,\ldots,n+1$ runs over the $O_{gg}$ double-traces of twist $4+2n$. We note that $M$ is orthogonal, since 
\es{}{
(MM^T)_{rs}=\sum_a M_{ra}M_{as}=\frac{\sum_a C^{(0)}_{rra}C^{(0)}_{ssa}}{\sqrt{\sum_b (C^{(0)}_{rrb})^2}\sqrt{\sum_c (C^{(0)}_{ssb})^2}}=\delta_{rs}\,,
}
which follows from the fact that GFT $\langle ppqq\rangle$ correlators only exchange the identity for $p\neq q$. As a result, we also have
\es{M^TM_gluons}{
(M^TM)_{ab}=\sum_p M_{ap}M_{pb}=\frac{\sum_p C^{(0)}_{ppa}C^{(0)}_{ppb}}{{\sum_c(C^{(0)}_{ppc})^2}}=\delta_{ab}\,.
}
Using  \eqref{M^TM_gluons} in \eqref{log2directPRE}, we find
\es{}{
F^{(s)}_{n,\ell}&=\frac{1}{N^3}
\left[C^{(0)}_{\bft\bft i}\Gamma_{ia}^{(\text{tree})}(M^TM)_{ab}\Gamma_{bc}^{(\text{tree})}C^{(0)}_{22 j}\right]_{n,\ell}\\
&=\frac{1}{N^3}\sum_p
\left[\left(C^{(0)}_{\bft\bft i}\Gamma_{ia}^{(\text{tree})}C^{(0)}_{pp a}\right)\left(\sum_b(C^{(0)}_{ppb})^2\right)^{-1}\left(C^{(0)}_{ppc}\Gamma_{cd}^{(\text{tree})}C^{(0)}_{22d}\right)\right]_{n,\ell}\,,
}
where the combinations of OPE coefficients and anomalous dimensions are the same appearing in the expansion of free theory and tree level correlators. In particular, we can extract
\es{AomegadefS}{
\left. \mathcal{G}_{pppp}^{(0)}(U,V) \right|_{\mathbf{1}}&=\sum_{n,\ell}A^{\mathbf{1},(pppp)}_{n,\ell}U^{-1}G_{2p+2+2n+\ell,\ell}(U,V)\,,\\
\left. \mathcal{G}_{22pp}^{(\text{tree})}(U,V)\right|_{\mathbf{1},\log U}&=\frac{1}{N}\sum_{n,\ell}\omega^{\mathbf{1},(22pp)}_{n,\ell}U^{-1}G_{2p+2+2n+\ell,\ell}(U,V)\,,\\
\left. \mathcal{G}_{\bft \bft pp}^{(\text{tree})}(U,V)\right|_{\mathbf{1},\log U}&=\frac{1}{N^2}\sum_{n,\ell}\omega^{\mathbf{1},(\bft \bft pp)}_{n,\ell}U^{-1}G_{2p+2+2n+\ell,\ell}(U,V)\,,
}
where
\es{}{
A^{\mathbf{1},(pppp)}_{n,\ell}&=\left[C^{(0)}_{pp a}C^{(0)}_{ppa}\right]_{n+p-2,\ell}\,,\\
\omega^{\mathbf{1},(22pp)}_{n,\ell}&=\left[C^{(0)}_{22 a}\Gamma_{ab}^{(1)}C^{(0)}_{ppb}\right]_{n+p-2,\ell}\,,\\
\omega^{\mathbf{1},(\bft\bft pp)}_{n,\ell}&=\left[C^{(0)}_{\bft\bft i}\Gamma_{ia}^{(1)}C^{(0)}_{pp a}\right]_{n+p-2,\ell}\,,
}
so that ultimately we can write
\es{Fsdef}{
F^{(s)}_{n,\ell}&=\sum_{p=2}^{n+2}\frac{\omega^{\mathbf{1},(22pp)}_{n-p+2,\ell}\,\omega^{\mathbf{1},(\bft\bft pp)}_{n-p+2,\ell}}{A^{\mathbf{1},(pppp)}_{n-p+2,\ell}}\,.
}
Note that the sub/superscripts $\mathbf{1}$ in the above equations are meant as a reminder that we are focusing on operators that are singlets of $SO(8)$ (as well as of $SU(2)_L\times SU(2)_R$).

\subsubsection{Crossed channel}

The story is similar in the crossed channel. Here we are interested in the exchange of long double trace operators in the adjoint of $SO(8)$, that are singlets of $SU(2)_R$ and have $SU(2)_L$ spin $j_L=1$. Operators of this type can be formed with either two gluons or one gluon and one graviton, and the degeneracy space at twist $\tau = 4+2n$ is spanned by
\es{}{
[\cO_2\cO_2]_n\,,\quad \ldots\,, \quad
[\cO_{n+2}\cO_{n+2}]_n\,,\quad 
[\cO_{2}\bfo_{\bft}]_n\,,\quad \ldots\,, \quad
[\cO_{n+2}\bfo_{\mathbf{n+2}}]_n\,.
}
Again, the matrix of norms is diagonal so we can just take it to be orthonormal. As before, for the free theory OPE coefficients we have
\es{}{
C^{(0)}_{ggA}=(C^{(0)}_{gga},0)\,,\quad
C^{(0)}_{g\bfg A}=(0,C^{(0)}_{g\bfg \alpha})\,,
}
where now the index $\alpha=1,\ldots,n+1$ runs over the $O_{g\bfg}$ double trace operators. For the anomalous dimensions matrix, we now have the decomposition
\es{}{
\Gamma_{AB}=
\begin{pmatrix}
\Gamma_{ab} & \Gamma_{a\beta}\\
\Gamma_{\alpha b}& \Gamma_{\alpha\beta}
\end{pmatrix}\,,
}
where the leading order contribution comes from exchange Witten diagrams in four-point functions as above. In particular, the key new ingredient is $\Gamma^{\alpha b}$ and we can extract its leading order behavior from
\es{}{
\left. \langle \bfg g g g\rangle^{(\text{tree})}
\right|_{\log U} \rightarrow C^{(0)}_{\bfg g A}\Gamma^{(\text{tree})}_{AB}C^{(0)}_{g g B}
=
C^{(0)}_{\bfg g \alpha}\Gamma^{(\text{tree})}_{\alpha b}C^{(0)}_{g g b}\,.
}
From this we read off the large $N$ behavior
\es{}{
\Gamma^{(\text{tree})}_{\alpha b}\sim O(1/N^{3/2})\,.
}
The quantity we want to compute is 
\es{Glooplog2V}{
\left. \mathcal{G}^{(R|F^2)}(U,V)\right|_{\log^2 V}=\frac{U^2}{V^2}\sum_{n,\ell}F^{(t)}_{n,\ell}V^{-1}G_{6+2n+\ell,\ell}(V,U)\,,
}
where using similar logic to the previous subsection we find
\es{Ftdef}{
F^{(t)}_{n,\ell}=\frac{1}{N^3}\sum_{p=2}^{n+2}\frac{(\omega^{\mathbf{28},j_L=1 ,(\bft 2 p p)}_{n-p+2,\ell})^2}{A^{\mathbf{28},j_L=1 ,(pppp)}_{n-p+2,\ell}}\,,
}
in terms of
\es{expansioncrossed}{
\left. \mathcal{G}_{pppp}^{(0)} \right|_{\mathbf{28},j_L=1}&=\sum_{n,\ell}A^{\mathbf{28},j_L=1,(pppp)}_{n,\ell}U^{-1}G_{2p+2+2n+\ell,\ell}(U,V)\,,\\
\left. \mathcal{G}_{\bft 2 pp}^{(\text{tree})}\right|_{\mathbf{28} ,j_L=1,\log U}&=\sum_{n,\ell}\omega^{\mathbf{28},j_L=1 ,(\bft 2 p p)}_{n,\ell}U^{-1}G_{2p+2+2n+\ell,\ell}(U,V)\,,
}
where we have emphasized that we are interested in the contribution from the adjoint repsentation of $SO(8)$ and spin $j_L=1$ of $SU(2)_L$.

\subsection{One loop}

We are now ready to assemble the ingredients discussed in the previous subsection to compute the one loop result $\mathcal{G}^{(R|F^2)}$. Let us start from terms proportional to $\log^2U$ in the direct channel OPE, which are computed using \eqref{Glooplog2U}, where the coefficients $F^{(s)}_{n,\ell}$ are defined in \eqref{Fsdef}. Two out of three of the ingredients appearing in \eqref{Fsdef} were already computed in \cite{Alday:2021ajh}, namely the GFT OPE coefficients from $\langle pppp\rangle$
\es{}{
A^{\mathbf{1},(pppp)}_{n,\ell}=&
\frac{\pi  (-1)^{\ell } (\ell +1) (n+1)_{p-2} (n+p+1)_{p-2} 2^{-4 (n+p)-\ell +1} (2 n+2 p+\ell ) \Gamma (n+p) }{7 \Gamma (p)^4 \Gamma \left(n+p-\frac{1}{2}\right) \Gamma \left(n+p+\ell +\frac{1}{2}\right)}\\
&\times (n+\ell +2)_{p-2} (n+p+\ell +2)_{p-2} \Gamma (n+p+\ell +1)\,,
}
and the averaged anomalous dimensions from $\langle 22pp\rangle$
\es{}{
\omega^{\mathbf{1},(22pp)}_{n,\ell}=\frac{3 \pi  (-1)^{p+1} \left(2^{-4 p-4 n-\ell+3} (n+1)_{p-1} \Gamma (2 p+n-1) \Gamma (p+n+\ell+1)\right)}{14 \Gamma (p) \Gamma (p-1) \Gamma \left(p+n-\frac{1}{2}\right) \Gamma \left(p+n+\ell+\frac{1}{2}\right)}\,,
}
where note that in both cases we have considered the projection on $SO(8)$ singlets. \footnote{In this case the normalization of the $SO(8)$ projectors is not relevant since it would cancel between the numerator and denominator in \eqref{Fsdef}, but as we shall see it will play a role in the determination of \eqref{Ftdef}.} The remaining ingredient is the averaged tree anomalous dimensions from $\langle \bft\bft pp\rangle$, which can be extracted from \eqref{Mtree_GGgg_finalN} (with the normalizations \eqref{normsGGgg22pp}) projecting on $SU(2)_L$ singlets. Given that the correlator \eqref{Mtree_GGgg_finalN} receives two distinct contributions, one from a stress tensor exchange and one from the exchange of (generalized) current multiplets, we find it convenient to separate the two contributions as
\es{}{
\omega^{\mathbf{1},(\bft\bft pp)}_{n,\ell}=\omega^{(\mathcal{T})}_{n,\ell}{}^{\mathbf{1},(\bft\bft pp)}+\omega^{(\mathcal{J})}_{n,\ell}{}^{\mathbf{1},(\bft\bft pp)}\,,
}
and we find 
\es{}{
\omega^{(\mathcal{T})}_{n,\ell}{}^{\mathbf{1},(\bft\bft pp)}=\delta_{\ell,0}\frac{\pi  (-1)^p 2^{-4 n-4 p} (n+p)^4 (n+1)_{p-1} \Gamma (n+p) \Gamma (n+2 p)}{\Gamma (p-1) \Gamma (p) \Gamma \left(n+p+\frac{1}{2}\right) \Gamma \left(n+p+\frac{3}{2}\right)}\,,
}
where note that only spinless operators receive anomalous dimensions, and
\es{}{
\omega^{(\mathcal{J})}_{n,\ell}{}^{(\bft\bft pp)}=-\frac{(-1)^p 2^{1-\ell-4 n-4 p} \left(1+(-1)^\ell\right) p \pi  \Gamma (1+\ell+n+p) \Gamma (-1+n+2 p) (1+n)_{-1+p}}{3 \Gamma (-2+p) \Gamma (p) \Gamma \left(-\frac{1}{2}+n+p\right) \Gamma \left(\frac{1}{2}+\ell+n+p\right)}\,.
}
Correspondingly, we split the function $F^{(s)}_{n,\ell}$ as
\es{}{
F^{(s)}_{n,\ell}=F^{(s,\mathcal{T})}_{n,\ell}+F^{(s,\mathcal{J})}_{n,\ell}
}
where using the results above one finds
\es{Fst}{
F^{(s,\mathcal{T})}_{n,\ell}=-\delta_{l,0}\frac{\pi  2^{-4 n-10} (n+1)^2 (n+2)^5 \Gamma (n+4)^2}{\Gamma \left(n+\frac{5}{2}\right) \Gamma \left(n+\frac{7}{2}\right)}
}
and 
\es{Fsg}{
      F^{(s,\mathcal{J})}_{n,\ell}=&\frac{\pi  (-1)^\ell \left((-1)^\ell+1\right) (n+1) (n+2) 2^{-\ell-4 n-6} \Gamma (\ell+n+3) \Gamma (\ell+n+4)}{(\ell+1) (\ell+2 n+4) \Gamma \left(n+\frac{3}{2}\right) (n+2)_\ell \Gamma \left(\ell+n+\frac{5}{2}\right)}\\
      &\Big\{p^{(1)}_{n,\ell}
      \left[\log (4)-\psi ^{(0)}\left(1+\tfrac{\ell}{2}\right)+\psi ^{(0)}\left(\tfrac{1+\ell}{2}\right)\right]
      +\\&+p^{(2)}_{n,\ell}\left[\log (4)-\psi ^{(0)}\left(2+\tfrac{\ell}{2}\right)+\psi ^{(0)}\left(\tfrac{5+\ell}{2}\right)\right]
      +\\&+ p^{(3)}_{n,\ell}\left[\log (4)-\psi ^{(0)}\left(2+\tfrac{\ell}{2}+n\right)+\psi ^{(0)}\left(\tfrac{5+\ell}{2}+n\right)\right]\Big\}\,,
}
where the $p^{(i)}_{n,\ell}$ are polynomials in $n$ and $\ell$, whose explicit form we omit for simplicity, while $\psi^{(0)}(x)$ is the digamma function.

Correspondingly, from \eqref{Glooplog2U} we can write
\es{}{
\left. \mathcal{G}^{(R|F^2)}(U,V)\right|_{\log^2U}=\mathcal{G}^{(\mathcal{T})}(U,V)+\mathcal{G}^{(\mathcal{J})}(U,V)\,,
}
and using (\ref{Fst}-\ref{Fsg}) we can compute both terms in a small $z$, $\bar{z}$ expansion from the OPE. For each fixed power of $\bar{z}$ we can resum the power series in $z$ and converting that result into a power series in $U$ for finite $V$ we find
\es{GTddisc}{
\mathcal{G}^{(\mathcal{T})}(U,V)=\sum_{m=2}^{\infty}U^m\bigg[\frac{P^{(m)}(V)}{(V-1)^{2m+1}}+\frac{Q^{(m)}(V)}{(V-1)^{2m+1}}\log{V}\bigg]\,.
}
This can be compared with an Mellin space ansatz
\es{MT}{
\mathcal{M}^{(\mathcal{T})}=\sum_{m=2}^{\infty}\frac{c^\mathcal{T}_m}{s-2m}\,,
}
where the single poles in $s$ produce the necessary $\log^2U$ and matching the $\log{V}$ terms with \eqref{GTddisc} fixes $c^\mathcal{T}_m=24(m-1)$. While the expression \eqref{MT} is naively divergent, the the divergence can be absorbed into the $G^{(R^2F^2)}$ contact terms that appear at the same order in $1/N^2$. This allows one to derive the resummed expression
\es{loopMT}{
\mathcal{M}^{(\mathcal{T})}=6 (s-2) \psi ^{(0)}\left(2-\frac{s}{2}\right)+a_1 s+a_0\,,
}
where we have introduced two constants $a_0$ and $a_1$ parametrizing renormalization ambiguities. Note that for large $s$ (which is relevant for the flat space limit), we observe the behavior
\es{Mtflat}{
\mathcal{M}^{(\mathcal{T})}\simeq 6 s\log(-s)+a_1\,s\,,\quad (s\to\infty)\,,
}
so that the coefficient $a_1$ can be thought of as an energy/length scale regularizing the logarithm. Since this is precisely the role of one of the $G^{(R^2F^2)}$ contact terms that we will discuss in the next subsection, it is not necessary to explicitly include $a_1$.

The same procedure can be carried out to compute $\mathcal{G}^{(\mathcal{J})}$. In this case, though, the sum extends over all spins and the expressions are slightly more complicated. In particular, expanding again for small $z$, $\bar{z}$ and resumming the $z$ expansion for each fixed power of $\bar{z}$, we find
\es{GJexp}{
\mathcal{G}^{(\mathcal{J})}(U,V)=\!\sum_{m=3}^{\infty}\bar{z}^m\!\bigg[&\frac{p_1^{(m-1)}(z)}{z^m}\!+\!\frac{p_2^{(m-1)}(z)}{z^m}\!\log{(1\!-\!z)}\\
&+\frac{p_3^{(m-1)}(z)}{z^m}\!\log^2{(1\!-\!z)}\!+\!\frac{p_4^{(m-1)}(z)}{z^m}\mathrm{Li}_2(z)\bigg]\,.
}
Converting again to a power series in $U$ for finite $V$, we can compare that expression with the Mellin space ansatz
\es{Bans}{
\sum_{m,n=3}^{\infty}\frac{c^\mathcal{J}_{m,n}}{(s-2m)(t-2n)}+\sum_{m,n=3}^{\infty}\frac{c^\mathcal{J}_{m,n}}{(s-2m)(\tilde{u}-2n)}+\dots\,,
}
where the numerator is the same for the two sums as requires by crossing symmetry (amounting to $t\leftrightarrow \tilde{u}$), while the $\ldots$ denote possible terms containing single poles in $s$ or $t$ separately. Comparing the $\log^2U\log^2V$ terms in the spacetime expression \eqref{GJexp} from those obtained inverting the Mellin transform of the ansatz \eqref{Bans} fixes
\es{cJcoeff}{
c^\mathcal{J}_{m,n}=\frac{24 (m-2) (n-2) \left(4 m^2 n-6 m^2+4 m n^2-23 m n+23 m-6 n^2+23 n-19\right)}{5 (m+n-5) (m+n-4) (m+n-3) (m+n-2)}\,,
}
and moreover we find that, with these coefficients, the ansatz \eqref{Bans} is sufficient to reproduce the whole \eqref{GJexp} (not just the part proportional to $\log^2V$), so we can actually drop the $\ldots$ in \eqref{Bans} and define 
\es{Bdef}{
B(s,t)=\sum_{m,n=3}^{\infty}\frac{c^\mathcal{J}_{m,n}}{(s-2m)(t-2n)}\,,
}
with the coefficients given in \eqref{cJcoeff}. This expression can also be resummed, following a procedure analogous to that explained in \cite{Alday:2021vfb}. We obtain
\es{Bstresummed}{
B(s,t)=&\left(\psi ^{(1)}\left(2-\frac{s}{2}\right)+\psi ^{(1)}\left(2-\frac{t}{2}\right)-\left(\psi ^{(0)}\left(2-\frac{s}{2}\right)-\psi ^{(0)}\left(2-\frac{t}{2}\right)\right)^2\right) R_0(s,t)\\
&+\psi ^{(0)}\left(2-\frac{s}{2}\right) R_1(s,t)+\psi ^{(0)}\left(2-\frac{t}{2}\right) R_1(t,s)+R_2(s,t)\,.
}
where
\es{}{
R_0(s,t)=&\frac{3 (s-4) (t-4) \left(2 s^2 (t-3)+s \left(2 t^2-23 t+46\right)-6 t^2+46 t-76\right)}{5 (s+t-10) (s+t-8) (s+t-6) (s+t-4)}\\
R_1(s,t)=&-\frac{3 (s-4) \left(2 s^3+s^2 (8 t-49)+s \left(6 t^2-93 t+306\right)-22 t^2+218 t-520\right)}{5 (s+t-10) (s+t-8) (s+t-4)}\\
R_2(s,t)=&-\frac{3 \pi ^2 (-4+s) (-4+t) \left(-76+2 s^2 (-3+t)+46 t-6 t^2+s (46+t (-23+2 t))\right)}{5 (-10+s+t) (-8+s+t) (-6+s+t) (-4+s+t)}\\&+\frac{3 (452-68 t+s (-68+t (-11+2 s+2 t)))}{5
(-10+s+t) (-8+s+t)}\\ &+b_0+b_1 (s+ t)\,,
}
where $b_0$ and $b_1$ are new renormalization ambiguities whose presence is due to the fact that the sum \eqref{Bdef} is also divergent. We also observe that for large $s$ and $t$ (relevant for the flat space limit)
\es{Bstflat}{
B(s,t)\sim \frac{6}{5} \left(-\frac{\left(\pi ^2+\log ^2\left(\frac{s}{t}\right)\right) \left(s^2 t^2\right)}{(s+t)^3}-\frac{s^2 (s+3 t) \log (-s)+t^2 (3 s+t) \log (-t)}{(s+t)^2}+\frac{st}{s+t}\right)\,,
}
which has the same dependence on $s$ and $t$ as that of a ten-dimensional scalar box diagram -- see Appendix \ref{app:flatspace} and specifically eq. \eqref{Abox10d} for more details.

At this point, we have an expression in Mellin space that reproduces all terms proportional to $\log^2U$ in the small $U$ expansion, which we have computed solving a mixing problem following the results of the previous subsection. We now need to include new terms to also reproduce all terms proportional to $\log^2V$ for small $V$, see \eqref{Glooplog2V}. To this end, we make an ansatz for the one loop reduced Mellin amplitude as
\es{}{
\mathcal{M}^{(R|F^2)}&=
\mathcal{M}^{(\mathcal{T})}+\mathcal{M}^{(\mathcal{J})}\,,
}
where $\mathcal{M}^{(\mathcal{T})}$ was computed above while
\es{MJansall}{
\mathcal{M}^{(\mathcal{J})}=B(s,t)+B(s,\tilde{u})+\sum_{m,n=3}^{\infty}\frac{d^\mathcal{J}_{m,n}}{(s-2m)(t-2n)}+\dots\,,
}
and note that the only unknown part is the last sum, which we have included to account for terms proportional to $\log^2V$ in the crossed channel OPE (while note that it does not produce any $\log^2U$ term). The $\ldots$ account once again for separate single poles in $s$ and $t$. 

To compute the coefficients $d^\mathcal{J}_{m,n}$ we first assemble the various ingredients appearing in \eqref{Ftdef}. First, from the GFT correlator $\langle pppp\rangle$, we find \cite{Alday:2021ajh}\footnote{To be consistent with our conventions, we have to change the normalization of the projector on the $\mathbf{28}$, which is the adjoint of $SO(8)$. In particular, comparing with the definitions of \cite{Behan:2023fqq}, we have 
\es{}{
P_{\mathbf{28}}^{(\text{here})}=12 P_{\mathbf{28}}^{(\text{there})}\,,
}
where the $12=\psi^2h^\vee$, $\psi^2=2$ being the length squared of the longest root of the Lie algebra of $SO(8)$ and $h^\vee=6$ its dual Coxeter number.
}
\es{}{
      A^{\mathbf{28},j_L=1,(pppp)}_{n,\ell}=14\frac{p-2}{p}A^{\mathbf{1},(pppp)}_{n,\ell}\,,
      }
while from \eqref{MtreeGggg} we can extract
\es{}{
\omega^{\mathbf{28},j_L=1 ,(\bft 2 p p)}_{n,\ell}=  -\frac{(-1)^p 2^{3-\ell-4 n-4 p} \pi  \Gamma (n+p) \Gamma (1+\ell+n+p) \Gamma (-1+n+2 p)}{2 \sqrt{2} (p-3)!\Gamma (1+n) \Gamma (p) \Gamma \left(-\frac{1}{2}+n+p\right)
      \Gamma \left(\frac{1}{2}+\ell+n+p\right)}\,.
}
Carrying out a similar computation as that performed for $\log^2U$ terms, we find that this fixes
\es{}{
d^\mathcal{J}_{m,n}=c^\mathcal{J}_{m,n}\,,
}
and again the coincident poles in $s$ and $t$ in \eqref{MJansall} are sufficient to reproduce the full $\log^2V$ term, with no need for separate single poles, so we can drop the $\ldots$ like before here too.

We thus conclude that the reduced Mellin amplitude at one loop can be expressed as
\es{MRF2}{
\mathcal{M}^{(R|F^2)}&=
\mathcal{M}^{(\mathcal{T})}+\mathcal{M}^{(\mathcal{J})}\\
&=\mathcal{M}^{(\mathcal{T})}+B(s,t)+B(s,\tilde{u})+B(t,\tilde{u})\,,
}
where $\mathcal{M}^{(\mathcal{T})}$ is given in \eqref{loopMT} and $B(s,t)$ in \eqref{Bstresummed}. 

\subsection{Higher derivative corrections}

Finally, we consider higher-derivative corrections. As noted below \eqref{introG}, symmetries in principle allow for the presence of a contact term of the schematic type $RF^2$, which would correspond to a constant term (since it involves four derivatives)
\es{contact_vanishes}{
\frac{1}{N^{5/2}}\mathcal{M}^{(RF^2)}=\frac{a^{(RF^2)}(\tau)}{N^{5/2}}\,,
}
in the large $N$ expansion of the reduced Mellin amplitude. We will show in the next Section that the integrated constraint \eqref{intconstraint} actually implies that $a^{(RF^2)}(\tau)=0$.

Beyond this, the first correction that we expect comes from a coupling of the schematic type $R^2F^2$, which has a total of six derivatives. It corresponds to a polynomial of degree one for the associated Mellin amplitude. Recalling that crossing symmetry requires our results to be symmetric under the exchange of $t$ and $\tilde{u}$, and that $s+t+\tilde{u}=6$, we find that 
\es{M_HD}{
\mathcal{M}^{(R^2F^2)}(\tau)&=b^0(\tau)+b^1(\tau)\,s\,,\\
\mathcal{M}^{\log}&=b^0_{\log}+b^1_{\log}\,s\,,
}
which we can find only up to four coefficients (two functions of $\tau$ and two constants) which cannot be fixed by bootstrap considerations alone. The rest of the paper is devoted to explaining how these coefficients can be computed using supersymmetric localization and the flat space limit.

\section{Supersymmetric localization }\label{sec:loc}

In this section we compute the derivatives of the mass-deformed free energy of the theory which appear in \eqref{intconstraint}. We consider a mass deformation with four identical masses $\mu_i\equiv \mu$, $i=1,\ldots,4$, associated with the Cartan generators of $SO(8)$, for the fundamental hypermultiplets. Moreover, we also add a mass deformation parameter $m$ for the antisymmetric hypermultiplet, which is charged under $SU(2)_L$. The partition function $Z(\mu,m)=e^{-F(\mu,m)}$ of the theory is computed using supersymmetric localization in terms of a matrix model integral as \cite{Pestun:2007rz}
\es{Z_def}{
Z(\mu,m)=\int [dX]\,e^{{-\frac{8\pi^2}{g_\text{YM}^2}}\,\text{tr}X^2}\,|Z_{\text{1-loop}}(X,\mu,m)|^2\,|Z_{\text{inst}}(X,\mu,m,\tau_\text{UV})|^2\,,
}
where $X$ are $2N\times 2N$ matrices in the Lie algebra $\mathfrak{sp}(2N)$ with eigenvalues $\{\pm x_1,\ldots,\pm x_N\}$, and the Vandermonde measure is
\es{Van}{
[dX]=\frac{1}{N!}\prod_{i=1}^N dx_n\,x^2_n\prod_{1\leq n<m\leq N}(x^2_n-x^2_m)^2\,.
}
The other ingredients are the one-loop determinant 
\es{}{
|Z_{\text{1-loop}}(X,\mu,m)|^2=Z_{SO(8)}(\mu)Z_{SU(2)}(m)Z_{\text{int}}\,,
}
and the instanton partition function, to which we will get later. In the above we introduced 
\es{}{
Z_{SO(8)}(\mu)&=\frac{\prod_{i}H^8(x_i)}{\prod_{i}H^4(x_i+\mu)H^4(x_i-\mu)}\,,\\
Z_{SU(2)}(m)&=\frac{H^{N-1}(0)\prod_{i<j}H^2( x_{ij}^{-})H^2( x_{ij}^{+})}{H^{N-1}(m)\prod_{i<j}H( x_{ij}^{-}+m)H( x_{ij}^{+}+m)H( x_{ij}^{-}-m)H( x_{ij}^{+}-m)}\,,\\
Z_{\text{int}}&=\frac{\prod_{i}H^2(2x_i)}{\prod_iH^8(x_i)}\equiv \exp\{-{\textstyle\sum_{i=1}^N} S_{\text{int}}(x_i)\}=e^{-S_\text{int}}\,,
}
in terms of $H(x)=e^{-(1+\gamma)x^2}G(1+i x)G(1-ix)$ where $G$ is the Barnes $G$ function and $\gamma$ the Euler number. We have also introduced the combinations of eigenvalues $x^\pm_{ij}=x_i\pm x_j$.

Our goal will be to compute the quantity
\es{Fprimedef}{
F' \equiv -\left. \partial^2_\mu\partial^2_m F(\mu,m)\right|_{\mu=m=0}\,,
}
in the large $N$ limit. We begin by doing so for small Yang-Mills coupling $g^2_{\text{YM}}$, keeping the product $\lambda_{\text{UV}}=g^2_{\text{YM}}\,N$ fixed, which we refer to as the regime of perturbative string coupling. We will then move to the case of arbitrary $g^2_{\text{YM}}$, and we will derive $F'$ as a modular invariant function of the complexified string coupling $\tau$.

\subsection{Perturbative string coupling}

In the perturbative case, the instanton term in \eqref{Z_def} is exponentially suppressed and does not contribute to perturbative terms in the large $N$ expansion. We can therefore effectively set it to one in this section. To compute the rest, we find it convenient to introduce the following notation for expectation values in the $\mathcal{N}=4$ SYM matrix model
\es{doubleexp}{
\exmm{f(x_i)}=\frac{1}{Z_{\mathcal{N}=4}}\int d^N x\frac{1}{N!}\prod_{i=1}^{N}x_i^2\prod_{i<j}^{N}(x_i^2-x_j^2)^2e^{-\frac{16\pi^2}{g^2}\sum_{i=1}^{N}x_i^2}f(x_i)\,,
}
where the normalization term
\es{}{
Z_{\mathcal{N}=4}=e^{-F_{\cN =4}}=\int d ^N x\frac{1}{N!}\prod_{i=1}^{N}x_i^2\prod_{i<j}^{N}(x_i^2-x_j^2)^2e^{-\frac{16\pi^2}{g^2}\sum_{i=1}^{N}x_i^2}\,,
}
is the partition function of $\mathcal{N}=4$ SYM with $USp(2N)$ gauge group, whose expression as a function of $N$ can be found in \cite{Beccaria:2022kxy}. With this definition, we can write
\es{F' expmm}{
F' & =\frac{\exmm{Z_{int}\partial_\mu^2Z_{SO(8)}|_{\mu=0}\,\partial_m^2Z_{SU(2)}|_{m=0}}}{\exmm{Z_{int}}}-\frac{\exmm{Z_{int}\partial_\mu^2Z_{SO(8)}|_{\mu=0}}}{\exmm{Z_{int}}}\frac{\exmm{Z_{int}\partial_m^2Z_{SU(2)}|_{m=0}}}{\exmm{Z_{int}}}\,,
}
and expanding
\es{finite lambda largeN}{
F'(N,\lambda_\text{UV})=N\,\text{F}_1(\lambda_\text{UV})+\text{F}_2(\lambda_\text{UV})+\frac{1}{N}\text{F}_3(\lambda_\text{UV})+O(1/N^2)\,,
}
we wish to compute these first few terms in an expansion for large $\lambda_\text{UV}$. The details are quite technical and are discussed in Appendix \ref{app:matmodel}. Here we present the results, which read 
\es{F1result}{
\mathrm{F}_1&=16-\frac{32 \pi ^2}{\lambda_\text{UV}}-\sum _{n=1}^{\infty } \frac{128 n \Gamma \left(-\frac{1}{2}+n\right)^2 \Gamma \left(\frac{1}{2}+n\right) \zeta (1+2 n)}{\pi ^{3/2} \Gamma (n) \lambda
_{\text{UV}}^{n+\frac{1}{2}}}\\
&=16-\frac{32 \pi ^2}{\lambda_\text{UV} }-\frac{64 \zeta (3)}{\lambda_\text{UV} ^{3/2}}-\frac{48 \zeta (5)}{\lambda_\text{UV} ^{5/2}}-\frac{405 \zeta (7)}{2 \lambda_\text{UV} ^{7/2}}-\frac{7875 \zeta (9)}{4\lambda_\text{UV} ^{9/2}} +O(\lambda_\text{UV} ^{-11/2})\,,
}
which is the only result that we have computed to all orders in the large $\lambda_\text{UV}$ expansion. On the other hand, for the next order we only give the first few orders
\es{F2largelambda}{
\mathrm{F}_2=&-12 \log \left(\frac{\lambda_\text{UV} }{4 \pi ^2}\right)+c_{\mathrm{F}_2}-\frac{48 \log (2) \zeta (3)}{\pi ^2 \sqrt{\lambda _{\text{UV}}}}\\
&+\frac{4 \left(4 \pi ^2 \zeta (3)-15 \log (2) \zeta (5)\right)}{\pi ^2 \lambda_{\text{UV}}^{3/2}}+\frac{9 \left(32 \pi ^2 \zeta (5)-315 \log (2) \zeta (7)\right)}{8 \pi ^2 \lambda _{\text{UV}}^{5/2}}+O(\lambda_{\text{UV}}^{-7/2})\,,
}
which we have determined up to an integration constant $c_{\mathrm{F}_2}$ which we know numerically to be $c_{\mathrm{F}_2}\simeq-16\pm 0.1$, see Appendix \ref{app:matmodel} for details. Finally, we have also computed a few orders in the expansion of $\mathrm{F}_3$, which reads
\es{}{
  \mathrm{F}_3 = & \frac{6 \log (2) \lambda _{\text{UV}}}{\pi ^2} - \frac{\sqrt{\lambda _{\text{UV}}} \left(144 \zeta (3) \log ^2(2) + 8 \pi ^4 \right)}{24 \pi ^4} - \frac{31}{6}\\
  &+ \frac{3 \left(256 \pi ^2 \zeta (3) \log (2) - 480 \zeta (5) \log ^2 (2)\right)}{64 \pi ^4 \sqrt{\lambda _{\text{UV}}}} \\
  & - \frac{9 \left(288 \pi ^4 \zeta (3) + 25200 \zeta (7) \log ^2 (2) - 5120 \pi ^2 \zeta (5) \log (2)\right)}{1024 \pi ^4 \lambda _{\text{UV}}^{3/2}}+O(\lambda_{\text{UV}}^{-5/2})\,.
}

\subsection{Non-perturbative string coupling}

We now wish to move to the case of finite YM coupling, which in principle involves including instanton correction to our results. Since that is computationally involved, we take an indirect route that will lead to an educated guess for the expression of $F'$ as a function of the string coupling $\tau$, for the first few orders in the large $N$ expansion. We begin by recalling that the IR 't Hooft coupling $\lambda$, related by the UV coupling $\lambda_\text{UV}$ by \cite{Douglas:1996js,Hollands:2010xa}
\es{lambdaUV-IR}{
\frac{1}{\lambda}=\frac{1}{\lambda_\text{UV}}+\frac{\log 2}{2\pi^2 N}\,,
}
is the quantity that is directly related to the complexified string coupling $\tau=\tau_1+i\,\tau_2$. In particular, we have
\es{}{
\lambda=\frac{8\pi N}{\tau_2}\,.
}
We can then rewrite the expression of $F'$ in terms of $\tau_2$, which will lead us to a natural guess for the expression of $F'$ as a function of $\tau$, at least for the first few orders in the large $N$ expansion. We find
\es{Fprimetau2}{
F'=&16N -12\log N+c_{\mathrm{F}_2}-\frac{31}{6 N}+\frac{c_{\mathrm{F}_4}}{N^2} \\
&-4\pi \tau_2+12 \log \tau_2+O(\tau_2^{-2})\\
&+\frac{1}{N^{1/2}}\left[-\frac{2 \sqrt{2} \tau_2^{3/2} \zeta (3)}{\pi ^{3/2}}-\frac{2}{3} \sqrt{2} \pi^{1/2}\tau_2^{-1/2}+O(\tau_2^{-3/2})\right]\\
&+\frac{1}{N^{3/2}}\left[\frac{\tau_2^{3/2} \zeta (3)}{\sqrt{2} \pi^{3/2}}-\frac{3 \tau_2^{5/2} \zeta (5)}{8 \sqrt{2} \pi ^{5/2}}+O(\tau_2^{-1/2})\right]\\
&+\frac{1}{N^{5/2}}\left[-\frac{405 \tau ^{7/2} \zeta (7)}{2048 \sqrt{2} \pi ^{7/2}}+\frac{9 \tau ^{5/2} \zeta (5)}{32 \sqrt{2} \pi ^{5/2}}-\frac{81 \tau ^{3/2} \zeta (3)}{512 \sqrt{2} \pi ^{3/2}}+O(\tau_2^{1/2})\right]+O(N^{-7/2})\,,
}
where $c_{\mathrm{F}_2}$ is the constant we introduced earlier. On the other hand, while we have not computed $\mathrm{F}_4$, from the pattern in the $\mathrm{F}_j$ with $j<4$ we predict that it will contribute to $F'$ with a constant (which we have called $c_{\mathrm{F}_4}$) and terms of order $N^{-n-\frac{1}{2}}\tau_2^{n-\frac{3}{2}}$, which we account for with $O(\tau_2^{\sharp})$ in the last three lines of \eqref{Fprimetau2} and $O(N^{-7/2})$. From this expression, we can formulate a guess for the modular completion in terms of non-holomorphic Eisenstein series 
\es{Efunc}{
 E(s,\tau)=&\frac{2\zeta (2s)}{\pi^s}\tau_2^s+2\sqrt{\pi}\tau_2^{1-s}\frac{\Gamma{(s-1/2)}}{\pi^s\Gamma{(s)}}\zeta(2s-1)\\
 &+\frac{2\sqrt{\tau_2}}{\Gamma{(s)}}\sum_{k\neq 0}\abs{k}^{s-\frac{1}{2}}\sigma_{1-2s}(\abs{k})K_{s-\frac{1}{2}}(2\pi\tau_2\abs{k})e^{2\pi i k \tau_1}\,.
}
For example, we note that the combination appearing in the second line of \eqref{Fprimetau2} gives
\es{}{
12\left(\frac{\pi}{3}\tau_2-\log\tau_2\right)=12 E(1,\tau)|_{\text{pert.}}\,,
}
which is the perturbative terms corresponding to
\es{}{
E(1,\tau)\equiv-2\log(\sqrt{\tau_2}|\eta(\tau)|^2)\,,
}
which we have defined in terms of the Dedekind eta function\footnote{This formula is equivalent to the limit of $s\to1$ in \eqref{Efunc} after throwing out the pole at $s=1$.}
\es{}{
\eta(\tau)=q^{\tfrac{1}{12}}\prod_{n=1}^\infty (1-q^{2n})\,,\qquad q\equiv e^{i\pi\tau}\,.
}
Repeating the exercise at each order in the $1/N$ expansion, we obtain \eqref{Fprimetaufinal}.

While computing generic instanton corrections from the matrix model \eqref{Z_def} is a challenging task, it is possible to provide evidence in support of \eqref{Fprimetaufinal} by computing the first instanton correction to the $O(N^0)$ term $E(1,\tau)$. In particular, we recall from \cite{Behan:2023fqq} that the naive expression for the instanton partition function $Z_{\text{inst}}$ contains extra degrees of of freedom, which can be removed by replacing
\es{}{
Z_{\text{inst}}\rightarrow \frac{Z_{\text{inst}}}{Z_{\text{extra}}}\,,
}
where $Z_{\text{extra}}$ was computed in \cite{Behan:2023fqq} at the first few orders in the instanton expansion and is eigenvalue-independent. Including only terms that are relevant for the current computation, we have
\es{}{
Z_{\text{extra}}(\mu,m,q)=1+\frac{q}{8}[109-96\mu^2+48m^2\mu^2]+O(q^2,\mu^4,m^4)\,.
}
As a result, the first instanton correction to \eqref{Fprimetau2} at $O(N^0)$ comes from
\es{}{
-\left. \partial^2_\mu\partial^2_m |Z_{\text{extra}}|^2\right|_{\mu=m=0}=-24(q+\bar{q})+O(q^2,\bar{q}^2)\,,
}
which combines with the perturbative terms to form the combination
\es{}{
12\left(\frac{\pi}{3}\tau_2-\log\tau_2+2(q+\bar{q})\right)+O(q^2,\bar{q}^2)\,,
}
corresponding to $12E(1,\tau)$ including the first instanton correction.

\section{Gluon-graviton scattering in AdS$_5$}\label{sec:glugrascattering}

We are finally ready to combine all the ingredients to put constraints on the coefficients appearing in the expressions for the higher-derivative terms $\mathcal{M}^{(R^2F^2)}$ and $\mathcal{M}^{(\log)}$ in \eqref{M_HD}. To do so, we are going to combine the integrated constraint \eqref{intconstraint} with the expression for a gluon-graviton scattering amplitude in the flat space limit.

Let us recall here the expression of the $\langle \bft\bft 22\rangle$ correlator in Mellin space, for the reader's convenience. From the results of Section \ref{sec:correlators} we find
\es{MellinExp}{
\mathcal{M}=-\frac{1}{N^2}\frac{4}{s-2}+\frac{1}{N^3}\left[\mathcal{M}^{(\mathcal{T})}+\mathcal{M}^{(\mathcal{J})}+b^0(\tau)+b^1(\tau)\,s+(b^0_{\log}+b^1_{\log}\,s)\,\log N\right]\,,
}
see \eqref{MRF2} for the explicit expression of $\mathcal{M}^{(\mathcal{T})}$ and $\mathcal{M}^{(\mathcal{J})}$. We now apply the constraint \eqref{intconstraint} using the expression \eqref{Fprimetaufinal} for the left-hand side, and \eqref{intM} to perform the integral of each amplitude directly in Mellin space. At leading order, using $I[1/(s-2)]=-1/16$ we find agreement between the graviton exchange term in \eqref{MellinExp} and the leading order term in the expansion of $F'$. 

Next, we should address the contact term discussed in \eqref{contact_vanishes}, where we have claimed that it vanishes. This is easily proved using the integrated constraint, as the coefficient $a^{(RF^2)}(\tau)$ would be related to terms in the expansion of $F'$ in \eqref{Fprimetaufinal} appearing at $O(N^{1/2})$. Given the absence of such terms and the fact that $I[1]\neq0$, we conclude that $a^{(RF^2)}(\tau)=0$, which is consistent with the flat space expression of the amplitude \cite{Stieberger:2009hq} and with the results of \cite{Chester:2023qwo} for an analogous setup in M-theory.

Moving on, we look at the $\log N$ term in \eqref{Fprimetaufinal}, which can be used to obtain\footnote{Note that $I[1]=\frac{1}{24}$ and $I[s]=\frac{1}{12}$.}
\es{}{
2b^0_{\log}+4b^1_{\log}=-9\,.
}
Finally, note that we were able to determine the $O(N^0)$ term in \eqref{Fprimetaufinal} only up to a $\tau$-independent term $c_{\mathrm{F}_2}$, so we are only able to fix the $\tau$ dependence of the coefficients up to the addition of such a term. We find
\es{}{
2b^0_{\log}+4b^1_{\log}=-9E(1,\tau)+(\tau\text{-independent})\,.
}

\begin{figure}[h!]
\centering	
\includegraphics[scale=1.2]{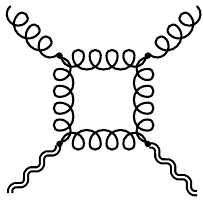}\hspace{4cm}
 \includegraphics[scale=1.1]{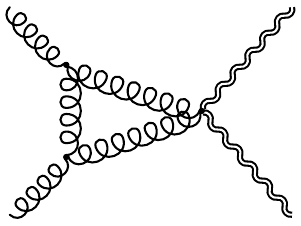}
\caption{The two types of Feynman diagrams contributing to the flat space limit \eqref{A_flat_fromM} of the one loop correlator \eqref{MRF2}.}
	\label{fig:loop}
\end{figure}

Note that we started with two pairs of coefficients and, naturally, since \eqref{intconstraint} provides a single constraint for each pair, we still have two free parameters. To fix those, we look at the flat space limit. Applying the formula \eqref{flatspacelim} to \eqref{MellinExp}, we find 
\es{A_flat_fromM}{
      \mathcal{A}(s, t)= & 8\pi^7(u+\alpha s )^2 \ell_P^8 \left[\frac{1}{s}  +\frac{\ell_P^{4}\pi}{20} (\mathcal{A}_{\text{box}}(s,t)\!+\!\mathcal{A}_{\text{box}}(s,u)\!+\!\mathcal{A}_{\text{box}}(t,u)-5 s \log (-s) )\right.\\
      & \left.\hspace{2.5cm}-\frac{\ell_P^{4}\pi}{24}b^1(\tau)s+\frac{\pi}{12}\ell_P^4\log{\ell_P^2} b^1_{\log}s\right]\,,
}
in terms of the Planck length $\ell_P=g_s^{1/4}\ell_s$. There are two types of contributions in \eqref{A_flat_fromM}: one is the crossing symmetric combination of the function $B(s,t)$, which we have computed using the gluon exchange term in \eqref{Mtree_GGgg_finalN}. In the flat space limit, this gives rise to a {\it ten}-dimensional scalar box diagram, as shown in \eqref{Bstflat}, that here we have called $\mathcal{A}_{\text{box}}$ -- see \eqref{Abox10d} for its precise definition. The other contribution comes from considering the graviton exchange term in \eqref{Mtree_GGgg_finalN} and in the flat space limit it reduces to the $s\log(-s)$ term in \eqref{A_flat_fromM}, which is an {\it eight}-dimensional scalar triangle diagram, as one can see from the $D\to 8$ limit of eq. \eqref{F3def}. There are two facts that should be noted about this. The first is that we have a combination of box and triangle diagrams, as represented visually in Figure \ref{fig:loop}. This is a novelty compared to the cases of maximal supergravity \cite{Alday:2017vkk,Alday:2020tgi,Alday:2021ymb,Alday:2022rly} or pure gluon scattering at one loop \cite{Alday:2021ajh,Behan:2022uqr} considered so far in the literature on holographic correlators, where only box diagrams are present.
The second observation is that in \eqref{A_flat_fromM} we see box diagrams in ten dimensions combined with a triangle diagram in eight dimension, both from taking the same ten-dimensional flat space limit using \eqref{flatspacelim}. This fact can be explained by a direct computation of the flat space scattering amplitude, which we perform in Appendix \ref{app:flatspace}. There, we show how the amplitude is written in terms of one loop box and scalar diagrams, where the loop momentum is always that of a gluon and therefore confined to live in {\it eight} dimensions. However, the special combination of eight-dimensional box diagrams appearing in the result is such that it corresponds to a {\it ten}-dimensional scalar box diagram -- see \eqref{G4def} for more details.

We can fix the coefficient $b^1_{\log}$ by requiring that the coefficient of the $\log\ell^2_P$ term is such as to combine with the logarithms arising from loop diagrams to make their arguments dimensionless. We find that this happens for $b^1_{\log}=-3$, which combined with the result from the integrated constraint gives
\es{}{
b^0_{\log}=\frac{3}{2}\,,\quad
b^1_{\log}=-3\,.
}
Moreover, we expect that the flat space amplitude should be a regular function of the string coupling $g_s$ and the string length $\ell_s$. We can use this as a constraint on $b^1(\tau)$ as follows. The $\log\tau_2$ term arising from the perturbative expansion of $b^1(\tau)$ should combine with the $\log\ell^2_P$ term in \eqref{A_flat_fromM} in such a way that, when $\ell_P$ is converted to $g_s$ and $\ell_s$, there are no $\log g_s$ terms, while $\log\ell_s$ combines with the logarithms of the Mandelstam variables in the same way as $\ell_P$ does. This fixes
\es{}{
b^1(\tau)=2\log\tau_2+\text{reg}(\tau)\,,
}
which can be completed to a modular-invariant function  by
\es{b1tau}{
b^1(\tau)=-3E(1,\tau)+\tau\text{-independent}.
}
As a result, combining this with the integrated constraint we also have
\es{}{
b^0(\tau)=\frac{3}{2}E(1,\tau)+\tau\text{-independent}\,.
}
Note that an argument in support of the fact that $b^1(\tau)$ should be given by $E(1,\tau)$ only (up to $\tau$-independent terms) is provided by \cite{Kiritsis:2000zi}, which computed the $R^2F^2$ coupling in this model using string dualities, finding precisely this result (up to $\tau$-independent terms, which were not computed in \cite{Kiritsis:2000zi} either). Moreover, we also note that fixing coefficients in this way the perturbative expansion of the genus zero term gives
\es{}{
\left.\mathcal{A}(s,t)\right|_{\text{genus 0}}=
8\pi^7(u+\alpha s )^2 \ell_P^8 \left[\frac{1}{s}+\frac{\zeta(2)}{4}\ell_s^4s+O(\ell_s^6)\right]\,,
}
where the ratio between the field theory term and the first string correction agrees with the results of \cite{Stieberger:2009hq} for the scattering amplitude between two closed strings and two open strings living on a D brane in superstring theory.

The final result for the reduced Mellin amplitude can then be expressed as 
\es{Mellin 1-loop Result}{
      \mathcal{M}=-\frac{4}{N^2}\frac{1}{s-2}
      +\frac{1}{N^3}\left[
      \mathcal{M}^{(\mathcal{T})}+\mathcal{M}^{(\mathcal{J})}
      -\frac{3}{2}(\log N+E(1,\tau))(2s-1)+c_1s+c_0\right]\,,
}
where $c_0$ and $c_1$ are $\tau$-independent coefficients which we were not able to determine. 

%The associated flat space amplitude can be expressed as
%\es{flat space Result}{
%      \mathcal{A}=&  (u+\alpha\,s )^2 \ell_P^8    \left[\frac{1}{s}  +\frac{\pi \ell_P^{4}}{20} \Big({\mathcal{A}}_{\text{box}}(s,t)\!+\!{\mathcal{A}}_{\text{box}}(s,u)\!+\!{\mathcal{A}}_{\text{box}}(t,u)\right.\\
%      &\left. \hspace{4.5cm}-5 s \log (-\ell_P^2 s ) +\frac{5}{6}(3E(1,\tau)-c_1)s\Big)\right]\,.
%}
%Note that the term $s\log(-s)$ has changed to $s\log(-\ell_P^2s)$ after choosing the correct value for $b^1_{\log}$. Moreover, the logarithms in $\mathcal{A}_{\text{box}}(s,t)$ can be made dimensionless without introducing extra $\log\ell_P$ terms.

\section{Conclusion}\label{sec:conclusion}

In this paper we studied mixed gluon-graviton scattering on D7 branes in AdS at large $N$ and finite $\tau$. We computed the one loop correction and matched it to the expected flat space result in \cite{Porkert:2022efy}. We also computed the mass derivatives $\partial_\mu^2\partial_m^2F(\mu,m)\vert_{\mu=m=0}$ at large $N$ and finite $\tau$ in terms of non-holomorphic Eisensteins series. We used the relation between this quantity and an integral of the holographic correlator to show that the $RF^2$ contact term vanishes, and we combined the constraint with the known $\tau$-dependence in the flat space limit \cite{Kiritsis:2000zi} to fix the $\tau$-dependence of the $R^2F^2$ contact term, whose coefficient is proportional to $E(1,\tau)$, as well as the logarithmic threshold term.

Looking ahead, we would like to also fix the $\tau$-independent parts of the $R^2F^2$ contact term. To do this we would need to compute the $\tau$-independent part of the flat space result in \cite{Kiritsis:2000zi}. We would also need to compute the $\tau$-independent contribution $c_{\mathrm{F}_2}$ to $\partial_\mu^2\partial_m^2F(\mu,m)\vert_{\mu=m=0}$, which we currently have a rough numerical estimate for, but would be nice to compute in closed form. This will likely require new technical methods of taking the large $\lambda$ expansion of the finite $\lambda$ results at that order in $1/N$.

It would also be interesting to guess a finite $N$ expression for $\partial_\mu^2\partial_m^2F(\mu,m)\vert_{\mu=m=0}$. This was accomplished for two mass derivatives of the sphere free energy in $\mathcal{N}=4$ SYM \cite{Dorigoni:2021guq,Dorigoni:2022zcr}, which at large $N$ and finite $\tau$ was also written in terms of non-holomorphic Eisenstein series. The difference in our case is that $E(1,\tau)$ appears, and we also have $\tau$-independent constants at every integer power of $1/N$. As mentioned before, our result seems to have the same form at large $N$ and finite $\tau$ as the integrated correlator of two gravitons and two giant graviton D3 branes in $AdS_5\times S^5$ \cite{Brown:2024tru}, which can perhaps be used to find a finite $N$ expression for both.

There are several generalizations to our results. In 4d, one could consider two different orbifolds of this theory as described in \cite{Ennes:2000fu}, where instead of the $SO(8)$ flavor group one now has $U(4)$ or $SO(4)\times SO(4)$. Some localization results for the mass deformed sphere free energy were considered for the $U(4)$ theory in \cite{Beccaria:2021ism,Billo:2024ftq}. One could also consider the other F-theory AdS setups where $\tau$ is fixed to a specific value, and the flavor group is now $SU(2),SU(3),E_6,E_7,E_8$ \cite{Fayyazuddin:1998fb}. The dual CFTs in these case are non-Lagrangian, but it might be possible to study them using their known Seiberg-Witten curves, as was done for gluon scattering in \cite{Behan:2024vwg}. One could also consider gluon-graviton scattering in other AdS$_{d+1}$/CFT$_{d}$ duals, namely $d=3,5,6$. The $d=3$ case was already discussed in \cite{Chester:2023qwo}, but the one loop term was not yet computed. It should be possible to fix the unique contact term contributing at the same order in $1/N$ as the one loop term, using the recent localization results of \cite{Kubo:2025dot}. Lastly, one can consider the planar all orders in $1/\lambda$ generalization of our result, as was done recently for the case of gluon scattering in \cite{Alday:2024yax}.

\section*{Acknowledgments} 
We thank Fernando Alday, Daniele Dorigoni, and Boris Pioline for useful discussions. SMC is supported by the Royal Society under the grant URF\textbackslash R1\textbackslash 221310 and the UK Engineering and Physical Sciences Research council grant number EP/Z000106/1. The work of DRP is supported by STFC DTP research studentship grant ST/Y509231/1.

\appendix

\section{Free theory}\label{app:free}

In this appendix we discuss how to obtain some of the free theory results used in the main text. We are interested in correlation functions between the superconformal primaries of the super gluon and super graviton 1/2-BPS multiplets. These are built from products of the scalar fields in the fundamental and antisymmetric hypermultiplets of the theory.

There are eight fundamental half-hypermultiplets, so we have a set of complex scalar fields
\es{}{
\phi^{\alpha,a,i}(x)\,,
}
where $\alpha\in \{1,2\}$ is an $SU(2)_R$ index, $a\in\{1,\ldots,2N\}$ is a fundamental index of $USp(2N)$ and $i\in\{1,\ldots,8\}$ is a fundamental index of $SO(8)$. These scalars are subject to the reality condition
\es{}{
[\phi^{\alpha,i,a}(x)]^*=\epsilon_{\alpha\beta}\Omega_{ab}\phi^{\beta,j,b}(x)\,.
}
We can contract the $SU(2)_R$ index with a polarization tensor $y^{\alpha}$ to define
\es{}{
\phi^{a,i}(x,y)=\epsilon_{\alpha\beta}\phi^{\alpha,i,a}(x)y^\beta\,,
}
and the two-point function of $\phi$ reads
\es{}{
\langle \phi^{i,a}(x_1,y_1)\phi^{j,b}(x_2,y_2)\rangle=\frac{y_1\cdot y_2}{x_{12}^2}\Omega^{ab}\delta^{ij}\,.
}

On the other hand we also have two half-hypermultiplets in the antisymmetric traceless representation of $USp(2N)$, so another set of complex scalars
\es{}{
\psi^{\alpha,\bar{\alpha},[ab]}(x)\,,
}
where $\bar{\alpha}\in \{1,2\}$ is an $SU(2)_L$ index and $\psi$ is $\Omega$-traceless:
\es{}{
\Omega_{ab}\psi^{\alpha,\bar{\alpha},ab}(x)=0\,.
}
We again have a reality condition
\es{}{
[\psi^{\alpha,\bar{\alpha},ab}(x)]^*=\epsilon_{\alpha\beta}\epsilon_{\bar\alpha\bar\beta}\psi^{\beta,\bar{\beta},ab}(x)\,,
}
and we contract these scalars with $SU(2)_R$ polarizations $y^\alpha$ as well as $SU(2)_L$ polarizations $\bar{y}^{\bar\alpha}$, defining
\es{}{
\psi^{ab}(x,y,\bar{y})=\epsilon_{\alpha\beta}\epsilon_{\bar\alpha\bar\beta}\psi^{\alpha,\bar{\alpha},ab}(x)y^\beta\bar{y}^{\bar\beta}\,.
}
Finally, the two-point function between such scalars is
\es{}{
\langle \psi^{ab}(x_1,y_1,\bar{y}_1)\psi^{cd}(x_2,y_2,\bar{y}_2)\rangle=\frac{(y_1\cdot y_2)(\bar{y}_1\cdot \bar{y}_2)}{x_{12}^2}\left(\Omega^{a[c|}\Omega^{b|d]}-\frac{1}{2N}\Omega^{ab}\Omega^{cd}\right)\,.
}

We are then ready to define the superprimaries of the supergluon and supergraviton multiplets. For the former, we set
\es{}{
\mathcal{O}^A_p(x,y,\bar{y})=\mathcal{N}_p\,\phi^{a_0,i}(x,y)\Omega_{a_0a_1}\psi^{a_1b_1}(x,y,\bar{y})\Omega_{b_1a_2}\ldots 
\psi^{a_{p-2}b_{p-2}}(x,y,\bar{y})\Omega_{b_{p-2}a_{p-1}}\phi^{a_{p-1},j}(x,y)T^A_{ij}\,,
}
where $T^A_{ij}$ are $SO(8)$ generators in the fundamental representations, for which we follow the same conventions as \cite{Behan:2023fqq}, while $\mathcal{N}_p$ is a normalization factor which is computed in such a way that the two-point functions are unit-normalized:
\es{}{
\langle
\mathcal{O}^A_p(x_1,y_1)
\mathcal{O}^B_q(x_2,y_2)
\rangle=(12)^p(\bar{y}_1\cdot \bar{y}_2)^{p-2}
\delta_{pq}\delta^{AB}\,,\quad 
(ij)\equiv \frac{(y_i\cdot y_j)}{x_{ij}^{2}}\,.
}

On the other hand, for the gravitons we set
\es{}{
\bfo_{\bfp}(x,y,\bar{y})=\mathbf{N}_{\bfp}\psi^{a_1b_1}(x,y,\bar{y})\Omega_{b_1a_2}\psi^{a_2b_2}(x,y,\bar{y})\Omega_{b_2a_3}\ldots\psi^{a_pb_p}(x,y,\bar{y})\Omega_{b_pa_1}\,,
}
where again $\mathbf{N}_{\bfp}$ is a normalization factor which sets
\es{}{
\langle
\bfo_{\bfp}(x_1,y_1,\bar{y}_1)
\bfo_{\bfq}(x_2,y_2,\bar{y}_2)
\rangle
=(12)^p(\bar{y}_1\cdot \bar{y}_2)^{p}\delta_{pq}\,.
}

After computing the two normalizations for a few values of $p$, we are able to evaluate certain three- and four-point functions. For three-point functions of supergluons we find
\es{}{
\langle \mathcal{O}^A_{p_1}(x_1,y_2,\bar{y}_2)
&\mathcal{O}^B_{p_2}(x_2,y_2,\bar{y}_2)
\mathcal{O}^C_{p_1}(x_3,y_3,\bar{y}_3)\rangle\\
&=C_{p_1p_2p_3}f^{ABC}(12)^{p_1+p_2-p_3}(23)^{p_2+p_3-p_1}(13)^{p_1+p_3-p_2}\\
&\times(\bar{y}_1\cdot\bar{y}_2)^{p_1+p_2-p_3-2}(\bar{y}_2\cdot\bar{y}_3)^{p_2+p_3-p_1-2}(\bar{y}_1\cdot\bar{y}_3)^{p_1+p_3-p_2-2}\,,
}
where $f^{ABC}$ are $SO(8)$ structure constants and we have checked for various values of $p_1,p_2,p_3$ that 
\es{}{
C_{p_1,p_2,p_3}=\frac{1}{\sqrt{2N}}+O(1/N^{3/2})\,,
}
where the leading order does not depend on the external weights. We observe that the leading order result (first derived in \cite{Alday:2021odx} from bootstrap considerations) is exact when at least one of the three operators is the moment map ($p=2$), while it receives higher order corrections in general. Other interesting three-point functions for us are
\es{}{
\langle \bfo_{\bft}(1)\co_p(2)^A\co_p(3)^B\rangle=C_{\bft,p,p}\delta^{AB}(12)(13)(23)^{p-1}(\bar{y}_1\cdot\bar{y}_2)(\bar{y}_1\cdot\bar{y}_3)(\bar{y}_2\cdot\bar{y}_3)^{p-3}\,,
}
where we find that
\es{}{
C_{\bft,p,p}=\frac{p-2}{\sqrt{2k_{SU(2)_L}}}\,.
}

We can also compute the free theory correlators $\langle \bft 2pp\rangle$
\es{}{
G_{\bft 2pp}^{(\text{free})}(U,V,\alpha)=\frac{2(p-2)}{\sqrt{k_{SO(8)}k_{SU(2)_L}}}\frac{U^2}{V}\alpha(\alpha-1)\,,
}
which at leading order in the large $N$ expansion becomes
\es{Gggg_free}{
G_{\bft 2pp}^{(\text{free})}(U,V,\alpha)=\frac{(p-2)}{\sqrt{2}N^{3/2}}\frac{U^2}{V}\alpha(\alpha-1)+O(1/N^{5/2})\,.
}

Last, we look at the correlator $\langle \bft \bft pp\rangle$, for which we recall that there are three functions corresponding to the three $SU(2)_L$ representations in the twofold tensor product of the adjoint representation. The result reads (in the conventions of \eqref{GGgg_22pp})
\es{free_22pp_GGgg}{
G_{\bft \bft pp}^{(12),(\text{free})}(U,V,\alpha)&=1+\frac{(p-2)U}{2N^2V}[1+\alpha(V-1)+(p-3)\alpha(1-\alpha)U]+O(1/N^3)\,,\\
G_{\bft \bft pp}^{(13),(\text{free})}(U,V,\alpha)&=
\frac{U}{2N^2V}[-(p-2) (1-\alpha(1+V))+(p-2)(p-3)\alpha(\alpha-1)U\\
&\hspace{3cm}+2(p-3)^2\alpha^2 UV]+O(1/N^3)\,,\\
G_{\bft \bft pp}^{(14),(\text{free})}(U,V,\alpha)&=
\frac{U}{2N^2V^2}[(p-2) V(1-\alpha(V+1))+(p-2)(p-3)\alpha(\alpha-1)UV\\
&\hspace{3cm}+2(p-3)^2(1-\alpha)^2U]+O(1/N^3)\,.
}

\section{Matrix model computations}\label{app:matmodel}

In this appendix we give more details on the matrix model computations sketched in Section \ref{sec:loc}. Our goal is to compute the expression of $F'$ defined in \eqref{Fprimedef} in a large $N$ expansion, first as a function of $\lambda_{\text{UV}}$ for small $g^2_{\text{YM}}$ and then for finite $g^2_{\text{YM}}$. Our starting point for this appendix is equation \eqref{F' expmm}, which expresses $F'$ in terms of expectation values in the $\mathcal{N}=4$ SYM matrix model. To compute the various expectation values in \eqref{F' expmm} it is convenient to introduce the following integral representations
\beq
\partial_\mu^2Z_{SO(8)}|_{\mu=0}=\int d\omega \frac{16\omega}{\sinh^2\!\omega}\tr \sin^2(\omega X)=\int d\omega \frac{8\omega}{\sinh^2\!\omega}S_m(\omega),\eeq
\begin{align}
\partial_m^2Z_{SU(2)}|_{m=0}&=\int d\omega \frac{\omega}{\sinh^2\!\omega}\left(-4(\tr\sin^2(\omega X))^2+8N\tr \sin^2(\omega X)-2\tr\sin^2(2\omega X)\right)\\
&=\int d\omega \frac{\omega}{\sinh^2\!\omega}\left(-S_m^2(\omega)+4NS_m(\omega)-S_m(2\omega)\right)\,\nonumber
\end{align}
where we introduced the function
\beq
S_m(\omega)=2\tr\sin^2(\omega X)
\eeq
and note that $\tr X^n=2\sum_ix_i^n$. It is also convenient to introduce an analogous integral representation for $S_{\text{int}}$ as
\es{}{
      S_{\text{int}}(X)=\int \frac{4}{\omega \sinh^2\!\omega}\tr\, \sin^4(\omega X)\,.
}

These results can be replaced in the expression of $F'$ in \eqref{F' expmm} to simplify the computation of the expectation values. To obtain a large $N$ expansion, we should also express the result in terms of connected correlation functions, since the latter have a well-defined scaling at large $N$. This can be achieved using the identities
\es{}{
      \exmm{e^{A}}&=e^{\exmm{e^A}_c-1}\,,\\
      \exmm{e^{A}B}&=\exmm{e^{A}}\exmm{e^{A}B}_c\,,\\
      \exmm{e^{A}B_1 B_2}&=\exmm{e^{A}}\left[\exmm{e^{A}B_1B_2}_c+\exmm{e^{A}B_1}_c\exmm{e^{A}B_2}_2\right]\,,
}
and generalizations thereof, where $A$ and $B_i$ are single trace operators. We can then write
\es{}{
F'=8\int\frac{d w w}{\sinh^2\!w}\frac{d vv}{\sinh^2\!v}\big[(&4N-2\exmm{e^{-S_\text{int}}S_m(v)}_c)\exmm{e^{-S_\text{int}}S_m(w)S_m(v)}_c\\
&-\exmm{e^{-S_\text{int}}S_m(w)S_m^2(v)}_c-\exmm{e^{-S_\text{int}}S_m(w)S_m(2v)}_c\big]\,,
}
and expanding $e^{-S_{\text{int}}}$ in Taylor series we obtain an expression for $F'$ that can be used to derive its large $N$ expansion. This requires the computation of expectation values of the form 
\es{}{
\exmm{\tr \sin^2(\omega_1 X)\dots \tr \sin^2(\omega_n X) \tr \sin^4(\omega_{n+1} X)\dots \tr \sin^4(\omega_m X)}_c\,,
}
which we do introducing two more technical ingredients. One is the Laplace transform
\es{}{
      \mathcal{L}_{\omega\to s}[f(\omega)]=\int_{0}^{\infty}f(\omega)e^{-s \omega}d\omega\,,
}
which we use to compute, for instance, 
\es{}{
      \mathcal{L}_{\omega\to s}[\sin^2(\omega x)]=\frac{1}{2 s}+\frac{i}{4 (-i s+2 x)}-\frac{i}{4 (i s+2 x)}\,
}
and similarly for $\sin^4(\omega x)$, thus reducing the computation of expectation values of products between sine functions to a correlator of rational functions of the eigenvalues, which can be computed using the resolvents
\es{}{
      W^{(n)}(y_1,\dots,y_n)=N^{n-2}2^{-n}\exmm{\tr\big(\frac{1}{y_1-X}+\frac{1}{y_1+X}\big)\dots\tr\big(\frac{1}{y_n-X}+\frac{1}{y_n+X}\big)}_c\,.
}
Using standard techniques (see, {\it e.g.}, \cite{ekr15}), it is straightforward to derive the first few resolvents which read 
\es{}{
W^{(1)}(y)=&\frac{4 \left(y-\sqrt{y^2-\mu ^2}\right)}{\mu
^2}+\frac{-\frac{1}{y}+\frac{1}{\sqrt{y^2-\mu ^2}}}{2 N}+\frac{\mu
^4}{32 N^2 \left(y^2-\mu ^2\right)^{5/2}}+O(N^{-3})\,,\\
      W^{(2)}(y_1,y_2)=&\frac{2 y_1^2 y_2^2\!-\!\mu ^2 y_1^2\!-\!y_2^2 \mu ^2\!-\!2y_1 y_2 \sqrt{\!y_1^2\!-\!\mu ^2} \sqrt{\!y_2^2\!-\!\mu
^2}}{\left(y_1^2-y_2^2\right)^2 \sqrt{y_1^2-\mu ^2}
\sqrt{y_2^2-\mu ^2}}\\
&+\frac{\mu ^4}{8 N\! \left(y_1^2\!-\!\mu
^2\right)^{3/2} \!\left(y_2^2\!-\!\mu ^2\right)^{3/2}}+O(N^{-2})\,.\nonumber
}
It is then straightforward to take the inverse Laplace transform of the results obtained with this method, which results in expressions involving Bessel functions of the first kind. Finally, to obtain the expansions for large $\lambda_{\text{UV}}$ appearing in the main text, it is convenient to use the method involving the Mellin-Barnes representation of the Bessel functions already implemented in \cite{Behan:2023fqq}, see also \cite{Alday:2023pet}.
Following the method just described, we obtain the following integral representations. At first order we have the rather compact formula
\es{}{
      \mathrm{F}_1=2^6\!\!\int\!\!\frac{d w \,w^2}{\sinh^2\!w}\frac{dv \,v}{\sinh^2\!v}J_1\!\left(\!\frac{v\sqrt{\lambda_{\text{UV}}}}{\pi}\right) \chi(v,w)\,.
}
where
\es{}{
      \chi(x,y)=\frac{x J_0\!\left(\!\frac{x \sqrt{\lambda _{\text{UV}}}}{\pi }\right) J_1\!\left(\!\frac{y\sqrt{\lambda _{\text{UV}}}}{\pi }\right)-\!y J_1\!\left(\!\frac{x \sqrt{\lambda _{\text{UV}}}}{\pi }\right) J_0\!\left(\!\frac{y\sqrt{\lambda _{\text{UV}}}}{\pi }\right)}{y^2-x^2}\,.
}
However, already at the following order the expressions become more involved and are given in the attached Mathematica notebook.

We now briefly review how this method is applied to obtain the large $\lambda$ expansion of the integrals of interest. We can deal with the non-factorizable combination of Bessel functions, $\chi(v,w)$, by using the Bessel kernel identity
\es{}{
      \frac{\omega_1 J_0\left(\frac{\sqrt{\lambda} \omega_1}{\pi}\right) J_1\left(\frac{\sqrt{\lambda} \omega_2}{\pi}\right)-\omega_2 J_1\left(\frac{\sqrt{\lambda} \omega_1}{\pi}\right) J_0\left(\frac{\sqrt{\lambda} \omega_2}{\pi}\right)}{\omega_2^2-\omega_1^2}=\sum_{\ell=1}^{\infty} \frac{4 \ell \pi J_{2 \ell}\left(\frac{\sqrt{\lambda} \omega_1}{\pi}\right) J_{2 \ell}\left(\frac{\sqrt{\lambda} \omega_2}{\pi}\right)}{\omega_1 \omega_2 \sqrt{\lambda}}\,.
}
We thus obtain an integral of an infinite sum of products of Bessel functions, to which we can apply the Mellin representation of the Bessel functions
\es{}{
      J_\mu(x) J_\nu(x)=\int \frac{d s}{2 \pi i} \frac{\Gamma(-s) \Gamma(2 s+\mu+\nu+1)\left(\frac{x}{2}\right)^{\mu+\nu+2 s}}{\Gamma(s+\mu+1) \Gamma(s+\nu+1) \Gamma(s+\mu+\nu+1)}\,,
}
and
\es{}{
      J_\nu(x)=\int \frac{d t}{2 \pi i}\frac{\Gamma(-t)\left(\frac{1}{2} x\right)^{\nu+2 t}}{\Gamma(v+t+1)}\,.
}
This allows us to compute the integrals in $v$ and $w$ by applying the following formula 
\es{}{
      \zeta(2 n+1)=\frac{1}{4 \Gamma(2 n+2)} \int_0^{\infty} \frac{d \omega \,\omega^{2 n+1}}{\sinh ^2(\omega / 2)}\,.
}
The infinite sum over $\ell$ can be evaluated by first expanding the resulting expression at large $\ell$ and then, each order is trivially summed as $\sum_{\ell=1}^{\infty}l^{-s}=\zeta(s)$. We can now close the integration contour in $s$ and $t$ to get the asymptotic large $\lambda$ expansion. We note that for each order in $1/\lambda$, only a finite number of orders in the $1/\ell$ expansion contributes, thus making the $1/\ell$ expansion of the sum justified.

This procedure fails at order $O(N^0)$ because F${}_2$ can be written as
\es{}{
      \mathrm{F}_2=\mathrm{F}_2^{\text{easy}}-\mathrm{F}_2^{\text{hard}}
}
with 
\es{F2finite}{
      \mathrm{F}_2^{\text{hard}}\equiv  16\frac{\lambda_\text{UV}}{\pi^2}\!\!\int\!\!\frac{dw\,w^2}{\sinh^2\!w}\frac{dv\,v^3}{\sinh^2\!v}\frac{d \omega}{\sinh^2\! \omega}\chi(v,w)\left(2\chi(\omega,v)-\chi(2\omega,v)\right),
}
the presence of a nested triple integral complicates the derivation of the large $\lambda_\text{UV}$ expansion since using the Mellin-Barnes representation of the Bessel functions is no longer sufficient.\\
\cite{Alday:2023pet} discussed a similar integral where double infinite sum appears but the method developed is still limited to two integration variable and is not easly generalizable to our case.

However, we observe that taking a derivative with respect to $\sqrt{\lambda_\text{UV}}$ gives two contributions,  and each of them can be factorized as a product between a double integral and a single integral, both of which can be computed using the Mellin-Barnes technique already described. Integrating back in $\sqrt{\lambda_\text{UV}}$ we obtain the result for the original integral up to an integration constant, which we have called $c_{\mathrm{F}_2}$ in \eqref{F2largelambda}. This can be evaluated numerically by comparing the analytic result 
\es{}{
      \mathrm{F}_2^{\text{hard}}=&-\frac{8 \log (2) \lambda _{\text{UV}}}{\pi ^2}+\frac{16 \log (2) \sqrt{\lambda _{\text{UV}}}}{\pi ^2}+8 \log \left(\frac{\sqrt{\lambda _{\text{UV}}}}{\pi }\right)+c_{\mathrm{F}_2}^{\text{hard}}\\
      &-\frac{4      \left(2 \pi ^2-9 \log (2) \zeta (3)\right)}{\pi ^2 \sqrt{\lambda_{\text{UV}}}}+O(\lambda _{\text{UV}}^{-3/2})\,,
}
with the numerical evaluation of \eqref{F2finite}, we find $c_{\mathrm{F}_2}^{\text{hard}}=8.3\pm 0.1$, resulting in the determination for $c_{\mathrm{F}_2}$ in \eqref{F2largelambda}.

It would be challenging to significantly reduce the error in the determination of $c_{\mathrm{F}_2}$ because of two competing features of the integral in \eqref{F2finite}. On the one hand, the asymptotic expansion is accurate for large values of $\lambda_\text{UV}$, while for smaller values exponentially small contributions become important. On the other hand, the numerical evaluation of the triple integral becomes challenging for large values of $\lambda_\text{UV}$, as the integrand becomes highly oscillatory, resulting in large numerical errors. We found that the best trade off between these two effects is obtained for $\lambda_\text{UV}\sim 200-500$. At any rate, this problem is no longer present for $\mathrm{F}_{i>2}$, and all terms can be expanded for large $\lambda_\text{UV}$ using the Mellin-Barnes representation of the Bessel functions. However, the computation quickly becomes tedious and we have stopped at $\mathrm{F}_3$, which is sufficient to derive the expression \eqref{Fprimetaufinal} for $F'$ at finite $\tau$ up to $O(1/N^{5/2})$.

\section{Flat space amplitudes}\label{app:flatspace}

In this appendix we compute flat space scattering amplitudes with two external gluons and two external gravitons, focusing on the tree level exchange of a graviton (the flat space equivalent of $G^{(R)}$ in \eqref{introG}) and on a loop diagram with only gluons running in the loop (corresponding to $G^{(R|F^2)}$ in \eqref{introG}). Rather than attempting a direct computation, we use the double copy relations \cite{Bern:2010ue}. In particular, following \cite{Chiodaroli:2014xia,Chiodaroli:2017ngp,Porkert:2022efy} we compute scattering amplitudes in supersymmetric Einstein-Yang-Mills (EYM) theory from the product between YM amplitudes on one side and YM$+\phi^3$ theory on the other. The latter is a non supersymmetric theory describing the interaction between a YM gauge field with gauge group $\tilde{G}_1$ and a scalar field $\phi$ in the biadjoint representation of $\tilde{G}_1\times\tilde{G}_2$, where $\tilde{G}_2$ is a flavor group -- see \cite{Chiodaroli:2017ngp} for details. In particular, to obtain an amplitude between two gluons and two gravitons, we consider the product between a four-gluon amplitude in SYM theory and an amplitude between two gluons and two scalars in YM$+\phi^3$.

Note that we are not interested in the most general configuration of momenta and polarizations for the external fields, rather in the case where the polarizations of gluons and gravitons are transverse to the momenta of all particles involved. This simplifies the calculations and intermediate expressions, and in what follows we only reproduce results in this special kinematical configuration.

\subsection{Tree level}

The first amplitude that we are interested in is a graviton exchange in the $s$-channel at tree level, which appears at $O(\kappa^2)$. \footnote{Note that in principle the amplitude also includes the exchange of gluons in the $t$ and $u$ channels, but this contribution vanishes when the momenta of gluons and gravitons are orthogonal to their polarizations.} To compute it using the double copy relations, we consider first the gluon exchange amplitude between four external gluons in SYM theory, given for example in \cite{Behan:2023fqq}, which can be written as
\es{}{
\mathcal{A}^{\text{tree},\text{SYM}}_4=g^2\left(\frac{c_sn_s}{s}+\frac{c_tn_t}{t}+\frac{c_un_u}{u}\right)\,,
} 
where $g$ is the YM coupling constant, $c_{s,t,u}$ are the usual color factors associated with a gluon exchange diagram satisfying the Jacobi identity $c_s+c_t+c_u=0$ and the kinematic factors are given by
\es{}{
n_s&=-(\epsilon_1\cdot \epsilon_2)(\epsilon_3\cdot \epsilon_4)(u+\alpha s)\,,\\
n_t&=(\epsilon_1\cdot \epsilon_2)(\epsilon_3\cdot \epsilon_4)(1-\alpha)(u+\alpha s)\,,\\
n_u&=(\epsilon_1\cdot \epsilon_2)(\epsilon_3\cdot \epsilon_4)\alpha(u+\alpha s)\,,
}
where $\epsilon_i$ are the polarizations of the four external gluon and note that $n_s+n_t+n_u=0$. The variable $\alpha$ above is given by
\es{}{
\alpha^2=\frac{(\epsilon_1\cdot \epsilon_3)(\epsilon_2\cdot \epsilon_4)}{(\epsilon_1\cdot \epsilon_2)(\epsilon_3\cdot \epsilon_4)}\,,
} 
as in \cite{Alday:2021odx}, and is the analogue of the R-symmetry cross ratio $\alpha$ introduced in \eqref{crdef} after the flat space limit.

The second ingredient is the exchange of a scalar field $\phi$ between two gluons and two scalars in the YM$+\phi^3$ theory. A straightforward computation gives
\es{}{
\mathcal{A}_{2,2}^{\text{tree},\text{YM}+\phi^3}=\tilde{g}^2\text{tr}(T^AT^B)\left[\frac{\tilde{c}_s\tilde{n}_s}{s}+\frac{\tilde{c}_t\tilde{n}_t}{t}+\frac{\tilde{c}_u\tilde{n}_u}{u}\right]\,,
}
where $\tilde{g}$ is the YM coupling constant of the YM$+\phi^3$ theory, $\tilde{c}_{s,t,u}$ are color factors associated with the YM gauge group $\tilde{G}_1$ while $T^A,T^B$ are generators of the flavor group $\tilde{G}_2$. The kinematic factors in this case are given by
\es{}{
\tilde{n}_s=(\tilde{\epsilon}_1\cdot \tilde{\epsilon}_2)(u-t)\,,\quad
\tilde{n}_t=(\tilde{\epsilon}_1\cdot \tilde{\epsilon}_2)t\,,\quad
\tilde{n}_u=(\tilde{\epsilon}_1\cdot \tilde{\epsilon}_2)u\,,
}
where $\tilde{\epsilon}_i$ are the polarizations of the two external gluons and again $\tilde{n}_s+\tilde{n}_t+\tilde{n}_u=0$. 

Following the prescription of \cite{Chiodaroli:2017ngp}, the tree level scattering amplitude between two gravitons and two gluons in supersymmetric EYM theory is then given by
\es{}{
\mathcal{A}_{2,2}^{\text{tree},\text{EYM}}=\left(\frac{\kappa_{\text{10d}}}{4}\right)^2\left(\frac{n_s\tilde{n}_s}{s}+\frac{n_t\tilde{n}_t}{t}+\frac{n_u\tilde{n}_u}{u}\right)\,,
}
which using the results above gives
\es{Atree}{
\mathcal{A}_{2,2}^{\text{tree},\text{EYM}}=\frac{\kappa_{\text{10d}}^2}{8}(\zeta_1\cdot \zeta_2)(\epsilon_3\cdot (\epsilon_4)\frac{(u+\alpha s)^2}{s}=\frac{(2\pi)^7\ell_P^8}{16}(\zeta_1\cdot \zeta_2)(\epsilon_3\cdot (\epsilon_4)\frac{(u+\alpha s)^2}{s}\,,
}
where $\zeta_i$ are the polarization tensors of the two external gravitons, defined as $\zeta_i^{\mu\nu}=\epsilon_i^{(\mu}\tilde{\epsilon}_i^{\nu)}$ in terms of the underlying gluon polarization vectors.

\subsection{One loop}

The scattering amplitude between two gravitons and two gluons receives two contributions at one loop: one where only gluons run in the loop (at $O(\kappa_{\text{10d}}^2\mathtt{g}_{\text{8d}}^2)$) and one with both gluons and gravitons in the loop (at $O(\kappa_{\text{10d}}^4)$). We are going to focus on the former, which is the flat space equivalent of the one loop term we computed in the main text in Mellin space. The result was derived in eq. (5.17) of \cite{Porkert:2022efy} and reads
\es{}{
\mathcal{A}_{2,2}^{\text{1-loop},\text{EYM}}=6\kappa_{\text{10d}}^2\mathtt{g}_{\text{8d}}^2(u+\alpha s)^2(\epsilon_1\cdot \epsilon_2)(\epsilon_3\cdot \epsilon_4)\left[(J(1,2,3,4)+(1\leftrightarrow 2))+(3\leftrightarrow 4)\right]\,,
}
where in general dimension $D$
\es{}{
J(1,2,3,4)&=J_1+J_2+J_3\,,\\
J_1&=\int \frac{d^D\ell}{(2\pi)^D}\frac{(\tilde{\epsilon}_1\cdot \ell)(\tilde{\epsilon}_2\cdot \ell)}{\ell^2\ell^2_1\ell^2_{13}\ell^2_{134}}\,,\\
J_2&=\int \frac{d^D\ell}{(2\pi)^D}\frac{(\tilde{\epsilon}_1\cdot (\ell+k_3))(\tilde{\epsilon}_2\cdot \ell)}{2\ell^2\ell^2_3\ell^2_{31}\ell^2_{314}}\,,\\
J_3&=-\int \frac{d^D\ell}{(2\pi)^D}\frac{\tilde{\epsilon}_1\cdot \tilde{\epsilon}_2}{4\ell^2\ell^2_3\ell^2_{34}}\,,
}
where $\ell_{12\ldots n}=\ell+k_1+k_2+\ldots+k_n$, $k_i$ are the momenta of the external particles (1 and 2 being the gravitons, 3 and 4 the gluons) and $\epsilon_i$, $\tilde{\epsilon}_i$ the polarizations of the gluons in the underlying sYM and YM$+\phi^3$ theories, as for tree level. Note that the contributions from $J_1$ and $J_2$ are associated with box diagrams, while $J_3$ is a triangle diagram. This is represented schematically by the two diagrams in Figure \ref{fig:loop}.

The computation of $J_3$ is straightforward as it is a scalar integral and using standard techniques we obtain
\es{}{
J_3=-\frac{\tilde{\epsilon}_1\cdot \tilde{\epsilon}_2}{4}F_3^{(D)}(s)\,,
}
where
\es{F3def}{
F_3^{(D)}(s)=\int \frac{d^D\ell}{(2\pi)^D}\frac{1}{\ell^2\ell_1^2\ell_{12}^2}=-i\frac{2^{5-2D}\pi^{(3-D)/2}}{(D-4)\Gamma[\tfrac{D-3}{2}]\sin(\tfrac{\pi D}{2})}(-s)^{-3+D/2}\,,
}
is the expression for a triangle diagram in $D$ dimensions. On the other hand, $J_1$ and $J_2$ require the evaluation of tensor integrals where the numerator contains the loop momentum with free indices. Such integrals can be evaluated using the Passarino-Veltman (PV) reduction, which allows to express them in terms of scalar integrals \cite{Passarino:1978jh}. The three types of integrals we need to compute are
\es{}{
F_4^{(D)}(s,t)\equiv I_4=\int  \frac{d^D\ell}{(2\pi)^D}\frac{1}{\ell^2\ell_1^2\ell_{12}^2\ell_{123}^2}\,,\quad
I_4^\mu=\int  \frac{d^D\ell}{(2\pi)^D}\frac{\ell^\mu}{\ell^2\ell_1^2\ell_{12}^2\ell_{123}^2}\,,\quad
I_4^{\mu\nu}=\int \frac{d^D\ell}{(2\pi)^D} \frac{\ell^\mu\ell^\nu}{\ell^2\ell_1^2\ell_{12}^2\ell_{123}^2}\,.
}
For the first, we find \cite{Valtancoli:2011kr}
\es{}{
F_4^{(D)}(s,t)=&i\frac{4^{3-D}\pi^{(3-D)/2}}{(D-4)\Gamma[\tfrac{D-3}{2}]\sin(\tfrac{\pi D}{2})}\frac{1}{s^3t^3}\left[s^2(-t)^{D/2}\Re[{}_2F_1(1,\tfrac{D}{2}-2,\tfrac{D}{2}-1;1+t/s)] + (s\leftrightarrow t)\right]\,,
}
which is a $D$-dimensional scalar box diagram. For the second, using the PV reduction, we find
\es{}{
I_4^\mu=\frac{F_3^{(D)}(s)-F_3^{(D)}(t)}{u}(k_1+k_3)^\mu-\frac{F_4^{(D)}(s,t)}{2u}(u\,k_1-t\,k_2+s\,k_4)^\mu\,.
}
The last integral can be computed in a similar way, but the expression is more complicated. However, since $I_4^{\mu\nu}$ always appears contracted with gluon polarization vectors, in the special kinematic configuration that we consider all terms in $I_4^{\mu\nu}$ that are proportional to momenta with free indices do not contribute to the final result. We can then write
\es{}{
I_4^{\mu\nu}=G_4^{(D)}(s,t)\eta^{\mu\nu}+\ldots\,,
}
where the $\ldots$ denote terms which vanish after contraction with $\tilde{\epsilon}_i^\mu\tilde{\epsilon}_j^\nu$ in our special kinematics. Using the PV reduction, it is possible to compute
\es{G4def}{
G_4^{(D)}(s,t)=\frac{st\,F_4^{(D)}(s,t)-2sF_3^{(D)}(s,t)-2tF_3^{(D)}(t)}{4(D-3)u}=-2\pi\,F_4^{(D+2)}(s,t)\,,
}
where we have emphasized that the combination of $D$-dimensional scalar triangle and box diagrams in \eqref{G4def} reproduces a scalar box diagram in $D+2$ dimensions. The final result reads
\es{}{
\mathcal{A}_{2,2}^{\text{1-loop},\text{EYM}}=3&(2\pi)^{4}\ell_P^{12}(u+\alpha s)^2(\zeta_1\cdot \zeta_2)(\epsilon_3\cdot \epsilon_4)\\
&\times \left[F_3^{(D)}(s)+4\pi(F_4^{(D+2)}(s,t)+F_4^{(D+2)}(s,u)+F_4^{(D+2)}(t,u))\right]\,,
}
which is a crossing symmetric contribution of box diagrams plus a triangle diagram. We are now ready to take the limit $D\to 8$ and, neglecting a single pole at $D=8$ which leads to an undetermined renormalization constant, we find
\es{Aloop}{
\mathcal{A}_{2,2}^{\text{1-loop},\text{EYM}}=&\frac{(2\pi)^{8}}{640}\ell_P^{12}(u+\alpha s)^2(\zeta_1\cdot \zeta_2)(\epsilon_3\cdot \epsilon_4)\\
&\times \left[\mathcal{A}_{\text{box}}(s,t)+\mathcal{A}_{\text{box}}(s,u)+\mathcal{A}_{\text{box}}(t,u)-5s\,\log(-s)\right]\,,
}
where we have introduced the function
\es{Abox10d}{
\mathcal{A}_{\text{box}}(s,t)=\frac{s^2t^2}{(s+t)^3}(\pi^2+\log^2\tfrac{s}{t})+\frac{s^2(s+3t)\log(-s)+t^2(t+3s)\log(-t)}{(s+t)^2}-\frac{st}{s+t}\,,
}
which is a ten-dimensional scalar box diagram, up to the overall normalization and a renormalization constant.

We note that the combination of \eqref{Atree} and \eqref{Aloop} reproduces precisely the flat space limit of the sum of the graviton exchange term $\mathcal{M}^{(R)}$ in \eqref{MR} and the one loop term $\mathcal{M}^{(R|F^2)}$ in \eqref{MRF2}, using the ten-dimensional flat space limit formula \eqref{flatspacelim} -- see \eqref{A_flat_fromM}.

\bibliographystyle{JHEP}
\bibliography{GluonGraviton.bib}

\end{document}